\documentstyle[takayama,12pt]{article}   
\input{epsf}
\newcommand{\vsl}{v\hspace{-1.6mm}/}

\renewcommand{\theequation}{\arabic{section}.\arabic{equation}}

\begin{document}

\refNP
\final
                                      
\date{TIT/HEP-384/NP\\
      August, 1998}
                                      
\title{Hyperon Non-leptonic Weak Decays in the Chiral Perturbation
Theory I}

\author{K.~Takayama and M.~Oka
        \affiliation{Department of Physics, 
                    Tokyo Institute of Technology\\
             Meguro, Tokyo 152, Japan}  }

\maketitle

\abstract{
Hyperon non-leptonic weak decay amplitudes are studied in the chiral
perturbation theory. We employ the low energy effective weak
Hamiltonian which contains the perturbative QCD correction. To include
the non-perturbative QCD effect, quark currents of the effective
Hamiltonian are substituted with hadronic currents which
are color singlet and are derived by the chiral perturbation
theory. We find that the amplitudes caused by the product of hadronic
currents are small. It reproduce the small amplitudes of
$\Delta I=3/2$, which are derived by the strong interaction correction.
}

\thispagestyle{empty}
\newpage


\setcounter{equation}{0}
\setcounter{figure}{0}
\section{Introduction}

Hyperon non-leptonic decay amplitudes have been studied very well
experimentally\cite{Data}; but the corresponding theory is not yet
satisfactory\cite{Bijnens85,Donoghue86,Jenkins92b,Springer95}. The
difficulty comes from that the S- and P-wave amplitudes are not
reproduced simultaneously, and that the enhancement mechanism of the
$\Delta I=1/2$ amplitudes is not understood. It is expected that the
hyperon decay amplitudes reflect the internal structure of the hyperon
and, therefore, QCD corrections to the standard theory must play
important roles there. Especially it is difficult to take into
account the non-perturbative QCD effect. In order to estimate the
amplitudes, low energy effective theories of the hadron are used in
the previous study\cite{Donoghue86}.

Chiral perturbation theory is one of the effective theories for the
low energy hadron phenomena. The chiral perturbation theory has
systematic perturbation on the meson momentum, and is expected to
reproduce the low energy hadron phenomena fairly well. But it is known 
that the chiral perturbation theory cannot reproduce the hyperon
non-leptonic weak decay amplitudes. There may remain following
questions.
\begin{itemize}
  \begin{enumerate}
    \item What makes it impossible to reproduce the weak
               interaction in the chiral perturbation theory?
    \item Is the chiral perturbation theory enough to describe the
          strong interaction correction for the weak interaction?
  \end{enumerate}
\end{itemize}
Since the chiral perturbation theory is very useful, it will be
applied to more complicated hadron phenomena, for example
$Y+N\rightarrow N+N$. It is, therefore, very important to solve
above questions. 

To consider the role of the chiral perturbation theory, we derive the
correction of the weak interaction into two parts, the perturbative
QCD corrections and the non-perturbative QCD corrections. The
perturbative corrections are taken into account with the one-loop
correction to the standard theory, and the low energy effective weak
Hamiltonian is derived by using renormalization group method. The
effective weak Hamiltonian consists of the products of two quark
currents. The non-perturbative QCD corrections are introduced by the
chiral perturbation theory. The hyperon and its interaction with
pseudo-scalar meson are presented by the heavy baryon formalism of the 
chiral perturbation theory. In this study, quark currents in the
effective weak Hamiltonian are replaced by the hadron currents which
are derived from the chiral perturbation theory. The effective weak
Hamiltonian is, then, given in terms of the hadron operators.
Previously the chiral perturbation theory needs many low energy
constants to describe the hyperon non-leptonic weak
decay\cite{Borasoy98}. The parameter fitting with the experimental
data is the problem. But, in our method, low energy constants in the
effective weak Hamiltonian are derived from the Lagrangian in the
strong interaction. This effective weak Hamiltonian demand fewer
number of low energy constants.
And this Hamiltonian makes it possible to estimate the $\Delta I=3/2$
and the $\Delta I=1/2$ amplitudes separately.

The rest of the paper is organized as follows. Section 2 introduces
the effective weak Hamiltonian which includes the perturbative QCD
corrections. The effective Hamiltonian described by the hadron
operators is constructed in section 3. Section 4 presents the
numerical analysis and discussion on the hyperon weak decay
amplitudes. Section 5 concludes the paper with the comments on the
chiral perturbation theory and the hyperon non-leptonic weak decay
amplitudes.


\setcounter{equation}{0}
\setcounter{figure}{0}
\section{The Perturbative QCD Corrections to the Effective Weak
         Hamiltonian }

The hadronic weak interaction is described by the standard theory
where the weak bosons are exchanged between
quarks\cite{standard,weak}. The Hamiltonian
density of the weak interaction with quarks and weak gauge bosons
becomes
\begin{equation}
 {\cal H}_I(x)=\frac{g}{2\sqrt{2}}J^+_{\mu}(x)W^-_{\mu}(x)
               +\mbox{H.C.}
  \label{eqn:weak1}
\end{equation}
where $W^{\pm}$ is the charged $W$-boson field, $J^{\pm}_{\mu}$
is the charged left-handed current and $g$ is the coupling constant.
Eq.(\ref{eqn:weak1}) is applied to the hyperon non-leptonic weak
decay. Because the gauge bosons are very heavy, the Hamiltonian for
the non-leptonic decay is composed of four quark vertices, which include
the QCD correction on the weak vertex. The weak Hamiltonian for the
hyperon decay changes the strangeness. The weak transition matrix for
$|\Delta S|=1$ process is defined
by\cite{vzs77,Gilman79,Paschos90}
\begin{equation}
 \bigl<|{\cal H}^{\Delta S=1}_{eff}|\bigr>=
   -\frac{i}{2}\int d^4x
       \bigl<\mbox{Tr}{\cal H}_I(x){\cal H}_I(0)\bigr>
  \label{eqn:weak2}
\end{equation}
Using the operator product expansion, it is possible to write the 
effective Hamiltonian as a sum of four quark operators. The
coefficients of the operator depend on the mass scale. At the scale
$\mu=M_W$ the QCD running coupling constant $\alpha_s$ is small, the
coefficients can be expanded perturbatively in $\alpha_s$.
Paschos et al.\cite{Paschos90} deduced the following effective
Hamiltonian at $\mu=M_W$:
\begin{equation}
 {\cal H}^{\Delta S=1}_{eff}\left|_{\mu=M_W}\right.
  =\frac{G_f}{\sqrt{2}}\left[
        \xi_u\left(\bar{d}_{\alpha}u_{\alpha}\right)
             \left(\bar{u}_{\beta}s_{\beta}\right)
       +\xi_c\left(\bar{d}_{\alpha}c_{\alpha}\right)
             \left(\bar{c}_{\beta}s_{\beta}\right)\right]
      +{\cal H}^{top}_{pengin},
  \label{eqn:weak3}
\end{equation}
where $\alpha$, $\beta$ denote the color index. $\xi_q$ satisfies
$\xi_q \equiv V^*_{qd}V_{qs}$ where $V$ is the Cabibbo-Kobayashi-Maskawa 
matrix\cite{Kobayashi73}. The first term is the pure weak interaction
where the transferred momentum is lower than the $W$-boson mass
scale. This term has the $\Delta I=3/2$ component as well as the
$\Delta I=1/2$ component. The second term represents the one-loop QCD
correction caused by the top quark in the internal line, which is
called as penguin diagram. Since the top quark is heavier than the
$W$-boson, this second term is treated separately. The scale $\mu$ in
eq. (\ref{eqn:weak3}) is changed with the renormalization group method, 
and the effective Hamiltonian for the low mass scale are obtained,
where one-loop QCD correction are taken into account
perturbatively. In this calculation, the renormalization scale $\mu$
is changed to $\mu^2\simeq\mu^2_0$ where $\alpha (\mu^2_0)=1$ is
satisfied, as is prescribed in Bardeen et al.\cite{Bardeen87}.
The effective weak Hamiltonian becomes
\begin{equation}
  H^{\Delta S=1}_{eff}(\mu^2\approx \mu^2_0)=-\frac{G_f}{\sqrt{2}}
         \sum^6_{r=1}K_r O_r,
  \label{eqn:weak4}
\end{equation}
where the operators $O_r$ are given by
\begin{equation}
  \begin{array}{lcl}
   &&O_1 = (\bar{d}_{\alpha}s_{\alpha})_{V-A}
                  (\bar{u}_{\beta}u_{\beta})_{V-A} 
         -(\bar{u}_{\alpha}s_{\alpha})_{V-A}
                   (\bar{d}_{\beta}u_{\beta})_{V-A}
    \\
   &&O_2 = (\bar{d}_{\alpha}s_{\alpha})_{V-A}
                 (\bar{u}_{\beta}u_{\beta})_{V-A} 
         +(\bar{u}_{\alpha}s_{\alpha})_{V-A}
               (\bar{d}_{\beta}u_{\beta})_{V-A}
    \\
       &&\qquad\qquad +2(\bar{d}_{\alpha}s_{\alpha})_{V-A}
            (\bar{d}_{\beta}d_{\beta})_{V-A}
    +2(\bar{d}_{\alpha}s_{\alpha})_{V-A}
             (\bar{s}_{\beta}s_{\beta})_{V-A}
    \\
  &&O_3 \left( \Delta I=\frac{1}{2} \right) =\frac{1}{3}
      \left[(\bar{d}_{\alpha}s_{\alpha})_{V-A}
            (\bar{u}_{\beta}u_{\beta})_{V-A} 
           +(\bar{u}_{\alpha}s_{\alpha})_{V-A}
            (\bar{d}_{\beta}u_{\beta})_{V-A}\right.
    \\
       &&\qquad\qquad\qquad\qquad\qquad\left.
          +2(\bar{d}_{\alpha}s_{\alpha})_{V-A}
            (\bar{d}_{\beta}d_{\beta})_{V-A}
          -3(\bar{d}_{\alpha}s_{\alpha})_{V-A}
            (\bar{s}_{\beta}s_{\beta})_{V-A}\right]
    \\
  &&O_4 \left( \Delta I=\frac{3}{2}\right)=\frac{5}{3}
      \left[(\bar{d}_{\alpha}s_{\alpha})_{V-A}
            (\bar{u}_{\beta}u_{\beta})_{V-A} 
           +(\bar{u}_{\alpha}s_{\alpha})_{V-A}
            (\bar{d}_{\beta}u_{\beta})_{V-A}
           \right.
    \\
       &&\left.\qquad\qquad\qquad\qquad\qquad
           -(\bar{d}_{\alpha}s_{\alpha})_{V-A}
            (\bar{d}_{\beta}d_{\beta})_{V-A}
           \right]
    \\
   &&O_5 = (\bar{d}_{\alpha}s_{\alpha})_{V-A}
           (\bar{u}_{\beta}u_{\beta}+\bar{d}_{\beta}d_{\beta}
            +\bar{s}_{\beta}s_{\beta})_{V+A}
    \\
   &&O_6 = (\bar{d}_{\alpha}s_{\beta})_{V-A}
           (\bar{u}_{\beta}u_{\alpha}+\bar{d}_{\beta}d_{\alpha}
            +\bar{s}_{\beta}s_{\alpha})_{V+A}.
  \end{array}
  \label{eqn:operator1}
\end{equation}
The coefficients $K_r(r=1,\cdots,6)$ are obtained form the calculation 
of the Wilson coefficients and the Cabibbo-Kobayashi-Maskawa matrix
elements. Two types of the Wilson coefficients are adopted for the
comparison, whose conditions are $m_t=200\mbox{[GeV]}$,
$\mu_0=0.24\mbox{[GeV]}$, $\Lambda^{(4)}=0.10\mbox{[GeV]}$ and
$m_t=200\mbox{[GeV]}$, $\mu_0=0.71\mbox{[GeV]}$,
$\Lambda^{(4)}=0.316\mbox{[GeV]}$.
The coefficients are listed in Table~2.1. In both cases, $\mu_0$ is
defined so as to satisfy $\alpha_s(\mu^2)=1$. The operators are
described as a sum of products of the $SU(3)_f$ currents. Since each
current belongs to the $SU(3)_f$ octet representation, the effective
weak Hamiltonian belongs to the {\bf 8} or {\bf 27} dimensional
representation. In eq.(\ref{eqn:operator1}) the operator $O_3$ and
$O_4$ belong to the {\bf 27} representation. Only the operator $O_4$
has the $\Delta I=3/2$ component, and other operators have only
$\Delta I=1/2$ component. The Wilson coefficients in
eq.(\ref{eqn:weak4}) suggest the enhancement of the $\Delta I=1/2$
amplitudes, but it is known that this perturbative enhancement is not
enough\cite{Okun82}. The correction using the renormalization group
method does not contain full effect of the QCD correction to calculate 
the amplitudes of non-leptonic weak decays of hyperons. At the low
momentum, the soft gluons are exchanged between quarks, which causes
the non-perturbative effect. The estimation of the non-perturbative
effect is very important for the hyperon weak decay. It is, therefore, 
necessary to introduce a low energy effective model which includes
non-perturbative effects of QCD.


\setcounter{equation}{0}
\setcounter{figure}{0}
\section{Chiral Effective Weak Hamiltonian for Hyperon Decays}
\subsection{Chiral perturbation theory in heavy baryon formalism}

Chiral perturbation theory treats hadrons as elementary
fields\cite{Gasser84,Weinberg79}. The hadron properties can be
calculated by perturbative expansions with respect to hadron momenta,
quark masses and baryon mass
differences\cite{Jenkins91a,Jenkins91b}. The chiral perturbation is
valid if the momentum $k$ is sufficiently smaller than the chiral
symmetry breaking scale $\Lambda_{\chi}\sim 1\mbox{[GeV]}$. In the
same way the quark mass matrix $M=diag(m_u,m_d,m_s)$ is suppressed by 
a factor $m/\Lambda_{\chi}$.

The chiral Lagrangian is constructed with these expansions, requiring
the chiral invariance, Lorentz invariance and parity conservation. The 
lowest and next order chiral Lagrangians for meson fields are given by
\begin{eqnarray}
 {\cal L}_2 &=& \frac{f_{\pi}^2}{4}
    \mbox{Tr}\{({\cal D}_{\mu}\Sigma)({\cal D}^{\mu}\Sigma^{\dag})
    +\chi\Sigma^{\dag}+\Sigma\chi^{\dag} \},
  \label{eqn:mesonlag1}\\
   {\cal L}_4 &=& L_1(\mbox{Tr}\{{\cal D}_{\mu}\Sigma 
                        {\cal D}^{\mu}\Sigma^{\dag}\})^2
    +L_2(\mbox{Tr}\{{\cal D}_{\mu}\Sigma D_{\nu}\Sigma^{\dag}\}
         \mbox{Tr}\{{\cal D}^{\mu}\Sigma D^{\nu}\Sigma^{\dag}\})
    \nonumber\\
  &&+L_3\mbox{Tr}\{{\cal D}_{\mu}\Sigma {\cal D}^{\mu}\Sigma^{\dag}
          D_{\nu}\Sigma D^{\nu}\Sigma^{\dag}\}
      +L_4(\mbox{Tr}\{{\cal D}_{\mu}\Sigma {\cal D}^{\mu}\Sigma^{\dag}\}
       \mbox{Tr}\{\chi\Sigma^{\dag}+\Sigma\chi^{\dag} \})
    \nonumber\\
  &&+L_5\mbox{Tr}\{{\cal D}_{\mu}\Sigma {\cal D}^{\mu}\Sigma^{\dag}
         (\chi\Sigma^{\dag}+\Sigma\chi^{\dag})\}
      +L_6(\mbox{Tr}\{\chi\Sigma^{\dag}+\Sigma\chi^{\dag}\})^2
    \nonumber\\
  &&+L_7(\mbox{Tr}\{\chi\Sigma^{\dag}-\Sigma\chi^{\dag}\})^2
       +L_8\mbox{Tr}\{\Sigma\chi^{\dag}\Sigma\chi^{\dag}+
         \chi\Sigma^{\dag}\chi\Sigma^{\dag}\}
    \nonumber\\
  &&-iL_9\mbox{Tr}\{F^{\mu\nu}_R {\cal D}_{\mu}\Sigma D_{\nu} \Sigma^{\dag}
    +F^{\mu\nu}_L {\cal D}_{\mu}\Sigma^{\dag} D_{\nu} \Sigma \}
      +L_{10}\mbox{Tr}\{\Sigma F^{\mu\nu}_R \Sigma^{\dag} F^L_{\mu\nu}\}
    \nonumber\\
  &&+H_1\mbox{Tr}\{F^{\mu\nu}_R F^R_{\mu\nu}+F^{\mu\nu}_L F^L_{\mu\nu} \}
       +H_2\mbox{Tr}\{\chi\chi^{\dag}\},
  \label{eqn:mesonlag2}
\end{eqnarray}
where $L_1 \sim L_{10}$ are the coupling constants. These values are
determined phenomenologically and are shown in Table 3.1.
The meson field $\Sigma (x)$ is given by
\begin{equation}
  \Sigma (x)=\xi^2(x)=e^{2i\pi(x)/f_{\pi}},
    \qquad\qquad
  \xi(x)=e^{i\pi(x)/f_{\pi}}.
  \label{eqn:mesinf1}
\end{equation}
This field is transformed under $SU(3)_L\times SU(3)_R$ as
$\Sigma \longrightarrow L\Sigma R^{\dagger}$. In order to preserve the 
local chiral invariance, the external gauge fields ${\cal V}_{\mu}$
and ${\cal A}_{\mu}$ appear through covariant derivative of mesons
\begin{equation}
  \begin{array}{lcl}
  {\cal D}_{\mu}\Sigma &=& \partial_{\mu}\Sigma
      -i({\cal V}_{\mu}+{\cal A}_{\mu})\Sigma
      +i\Sigma({\cal V}_{\mu}-{\cal A}_{\mu})
  \\
  {\cal D}_{\mu}\Sigma^{\dag} &=& \partial_{\mu}\Sigma^{\dag}
      +i\Sigma^{\dag}({\cal V}_{\mu}+{\cal A}_{\mu})
      -i({\cal V}_{\mu}-{\cal A}_{\mu})\Sigma^{\dag},
  \end{array}
  \label{eqn:mesonf2}
\end{equation}
and through the field strength tensors
\begin{equation}
  \begin{array}{lcl}
  F^{\mu\nu}_L &=& \partial_{\mu}({\cal V}_{\mu}-{\cal A}_{\mu})
                  -\partial_{\nu}({\cal V}_{\mu}-{\cal A}_{\mu})
                 -i\left[({\cal V}_{\mu}-{\cal A}_{\mu}),
                         ({\cal V}_{\mu}-{\cal A}_{\mu}) \right]
   \\
  F^{\mu\nu}_R &=& \partial_{\mu}({\cal V}_{\mu}+{\cal A}_{\mu})
                  -\partial_{\nu}({\cal V}_{\mu}+{\cal A}_{\mu})
                 -i\left[({\cal V}_{\mu}+{\cal A}_{\mu}),
                         ({\cal V}_{\mu}+{\cal A}_{\mu}) \right].
  \end{array}
  \label{eqn:trans1}
\end{equation}
The external scalar and pseudo-scalar fields, ${\cal S}$ and ${\cal P}$ 
respectively, are introduced as $\chi=2B_0({\cal S}-i{\cal P})$ and
$\chi^{\dagger}=2B_0({\cal S}+i{\cal P})$, where the quark mass matrix 
$M$ is included in $S$. Under the $SU(3)_L\times SU(3)_R$ chiral
symmetry, these external fields have the following transformations,
\begin{equation}
  \begin{array}{lcl}
 &&({\cal S}+i{\cal P}) \longrightarrow R({\cal S}+i{\cal P})L^{\dag}
    \\
 &&({\cal S}-i{\cal P}) \longrightarrow L({\cal S}-i{\cal P})R^{\dag}
    \\
 &&\left({\cal V}_{\mu}-{\cal A}_{\mu}\right)
    \longrightarrow L\left({\cal V}_{\mu}-{\cal A}_{\mu}\right)L^{\dag}
      +iL(\partial_{\mu}L^{\dag})
    \\
 &&\left({\cal V}_{\mu}+{\cal A}_{\mu}\right) 
  \longrightarrow R\left({\cal V}_{\mu}+{\cal A}_{\mu}\right)R^{\dag}
      +iR(\partial_{\mu}R^{\dag}).
  \end{array}
  \label{eqn:trans2}
\end{equation}
We follow the heavy baryon formulation for the octet and decuplet
baryons. Introducing the four-velocity $v_{\mu}$, the momentum
$p_{\mu}$ of the baryon becomes\cite{Jenkins91a}
\begin{equation}
  p_{\mu}=m_B v_{\mu}+k_{\mu}
  \label{eqn:momentum}
\end{equation}
where $m_B$ is the average octet baryon mass and $k_{\mu}$ represents 
the residual off-shell momentum of the baryon interacting with
Goldstone bosons. In the heavy baryon formalism, the first term in
eq.(\ref{eqn:momentum}) is removed and momentum expansion can be
treated in the same way as the Goldstone bosons. The mass term
expansion gives power series of $1/m_B$.

Using the expansions, the lowest order chiral Lagrangian for baryon
fields is given by\cite{Jenkins91b}
\begin{eqnarray}
  {\cal L}^{(1)}_v &=& i\mbox{Tr}\bar{B}_v(v\cdot {\cal D})B_v
                   +2D\mbox{Tr}\bar{B}_v S^{\mu}_v\{A_{\mu},B_v\}
                   +2F\mbox{Tr}\bar{B}_v S^{\mu}_v[A_{\mu},B_v]
  \nonumber\\
  &&-i\bar{T}^{\mu}_v(v\cdot {\cal D})T_{v\mu}
      +{\cal C}\left(\bar{T}^{\mu}_v A_{\mu} B_v 
          +\bar{B}_v A_{\mu}T^{\mu}_v \right)
      +2{\cal H}\bar{T}^{\mu}_v S_{v\nu}A^{\nu}T_{v\mu}
  \label{eqn:baryonlag1}\\
  &&+\Delta m \bar{T}^{\mu}_v T_{v\mu},
  \nonumber
\end{eqnarray}
where $D$, $F$, ${\cal H}$ and ${\cal C}$ are the coupling constants
determined phenomenologically. The decuplet-octet mass difference is
$\Delta m=m_T-m_B$. The lowest order Lagrangian coupled to the scalar
and pseudo-scalar external fields is given by
\begin{equation}
  \begin{array}{lcl}
  {\cal L}^{(2)}_v &=&
   a_1\mbox{Tr}\bar{B}_v\left\{(\xi^{\dag}\chi\xi^{\dag}+\xi\chi^{\dag}\xi),
     B_v\right\}
 +a_2\mbox{Tr}\bar{B}_v\left[(\xi^{\dag}\chi\xi^{\dag}+\xi\chi^{\dag}\xi),
     B_v\right]
    \\
  &&+a_3\mbox{Tr}\bar{B}_vB_v\mbox{Tr}
      (\xi^{\dag}\chi\xi^{\dag}+\xi\chi^{\dag}\xi)
    \\
   &&+c_1\bar{T}^{\mu}_v(\xi^{\dag}\chi\xi^{\dag}+\xi\chi^{\dag}\xi)T_{v\mu}
   +c_2\bar{T}^{\mu}_vT_{v\mu}\mbox{Tr}
      (\xi^{\dag}\chi\xi^{\dag}+\xi\chi^{\dag}\xi).
  \end{array}
  \label{eqn:baryonlag2}
\end{equation}
The velocity dependent baryon fields are defined by
\begin{equation}
 \begin{array}{lcl}
  B_v(x) &=& \frac{1+\vsl }{2}e^{im_B\vsl v\cdot x}B(x),
  \\
  T^{\mu}_v(x) &=& \frac{1+\vsl}{2}e^{im_B\vsl v\cdot x}T^{\mu}(x),
 \end{array}
 \label{eqn:baryon1}
\end{equation}
where $B$ and $T$ correspond to the octet and decuplet baryon fields,
respectively. The decuplet baryon fields are represented by the
Rarita-Schwinger field $T^{abc}_{\mu}$ which contains both Lorentz
index $\mu$ and spinor indices $a$, $b$, $c$. This field satisfies a
constraint $v^{\mu}\cdot T^{abc}_{v\mu}=0$. The velocity dependent
field have the two component spinors which are derived from the four
component spinors by the projection. The Dirac gamma matrix
$\gamma^{\mu}$ of the baryon-pion couplings in the Lagrangian is
replaced by the velocity vector $v_{\mu}$ and $\gamma^{\mu}\gamma^5$
by the velocity dependent spinor operator
$S^{\mu}_v$\cite{Jenkins91a,Jenkins91b}.

The fields are transformed under $SU(3)_L\times SU(3)_R$ symmetry as
\begin{equation}
 \begin{array}{lcl}
  B &\longrightarrow& U(x)BU(x)^{\dag}\\
  T^{\mu}_{abc} &\longrightarrow& U^d_a U^e_b U^f_c T^{\mu}_{def},
 \end{array}
 \label{eqn:trans3}
\end{equation}
where $U(x)$ is defined by
$\xi(x)\longrightarrow L\xi(x)U^{\dagger}(x)=U(x)\xi(x)R^{\dagger}$.
The Lagrangian ${\cal L}^{(1)}$ represents the pions derivatively
coupled to the octet and decuplet baryons through the vector and
axial-vector fields,
\begin{equation}
  \bar{V}^{\mu}=\frac{1}{2}
    \left(\xi\partial^{\mu}\xi^{\dag}+\xi^{\dag}\partial^{\mu}\xi\right),
  \qquad\qquad
  \bar{A}^{\mu}=\frac{i}{2}
    \left(\xi\partial^{\mu}\xi^{\dag}-\xi^{\dag}\partial^{\mu}\xi\right).
  \label{eqn:current1}
\end{equation}
In order to preserve the local chiral invariance, the external gauge
fields ${\cal V}^{\mu}$ appears through covariant derivative
\begin{equation}
  \begin{array}{lcl}
  {\cal D}^{\mu}B_v &=& \partial^{\mu}B_v
                      +\left[ V^{\mu},B_v \right]\\
  {\cal D}^{\mu}T^{\nu}_v &=& \partial^{\mu}T^{\nu}_{abc}
          +\left( V^{\mu}\right)^d_a T^{\nu}_{dbc}
          +\left( V^{\mu}\right)^d_b T^{\nu}_{adc}
          +\left( V^{\mu}\right)^d_c T^{\nu}_{abd},
  \end{array}
\end{equation}
where $V^{\mu}$ is defined by
\begin{equation}
   V^{\mu} \equiv \bar{V}^{\mu}
  -\frac{i}{2}\xi^{\dag}\left({\cal V}^{\mu}-{\cal A}^{\mu}\right)\xi 
  -\frac{i}{2}\xi\left({\cal V}^{\mu}+{\cal A}^{\mu}\right)\xi^{\dag}.
  \label{eqn:trans5}
\end{equation}
And the external gauge field ${\cal A}^{\mu}$ appears through the
interaction term
\begin{equation}
  \begin{array}{lcl}
  {\cal L}^{(1)}_{v\;int}
   &=&2D\mbox{Tr}\bar{B}_vS^{\mu}_v\left\{A_{\mu},B_v\right\}
     +2F\mbox{Tr}\bar{B}_vS^{\mu}_v\left[A_{\mu},B_v\right]
   \\
   &&+{\cal C}\left(\bar{T}^{\mu}_vA_{\mu}B_v
                    +\bar{B}_vA_{\mu}T^{\mu}_v\right)
     +2{\cal H}\bar{T}^{\mu}_vS_{v\nu}A^{\nu}T_{v\mu}
  \end{array}
\end{equation}
where $A^{\mu}$ is defined by
\begin{equation}
   A^{\mu} \equiv \bar{A}^{\mu}
  -\frac{i}{2}\xi^{\dag}\left({\cal V}^{\mu}-{\cal A}^{\mu}\right)\xi 
  -\frac{i}{2}\xi\left({\cal V}^{\mu}+{\cal A}^{\mu}\right)\xi^{\dag}.
  \label{eqn:trans6}
\end{equation}
We apply the chiral Lagrangian
(\ref{eqn:mesonlag1}), (\ref{eqn:mesonlag2}), (\ref{eqn:baryonlag1})
and (\ref{eqn:baryonlag2}) to derive the effective weak Hamiltonian
for the hyperon decay.

\subsection{Chiral effective Hamiltonian for weak interaction}

The low energy effective weak Hamiltonian (\ref{eqn:weak4}) is given
as a sum of products of two quark currents. Hence, it is natural to
construct an effective hadronic weak Hamiltonian by substituting the
quark currents by hadron currents, term by term. But the quark
operators can not be replaced by the hadronic operators directly,
since the hadronic operators do not have the same property as the
quark operators. As the effective weak Hamiltonian includes all
the hyperon decay processes, we,therefore, introduce the following
three ansatz.
\begin{itemize}
  \begin{enumerate}
    \item Fierz transformation is applied to the four quark operators, 
          and the form of the four quark operators are changed.
    \item The effective weak Hamiltonian is constructed by summing up
          all the weak operators which are derived by the Fierz
          transformation.
    \item The quark currents are replaced by the hadronic currents which
          have the same symmetry under the chiral transformation.
  \end{enumerate}
\end{itemize}
For the first ansatz, the Fierz transformation is applied to the quark 
currents and they become
\begin{equation}
  \begin{array}{lcl}
    \left(\bar{d}_{\alpha}u_{\alpha}\right)_{V-A} 
    \left(\bar{u}_{\beta}s_{\beta}\right)_{V-A}
    &\longrightarrow& 
    \frac{1}{3}\left(\bar{u}_{\alpha}u_{\alpha}\right)_{V-A} 
    \left(\bar{d}_{\beta}s_{\beta}\right)_{V-A}
    \\
    \left(\bar{d}_{\alpha}s_{\beta}\right)_{V-A} 
    \left(\bar{u}_{\beta}u_{\alpha}\right)_{V+A}
    &\longrightarrow& 
    -\frac{32}{9}\left(\bar{d}_{\alpha}u_{\alpha}\right)_{S-iP} 
    \left(\bar{u}_{\beta}s_{\beta}\right)_{S+iP}.
  \end{array}
  \label{eqn:trans7}
\end{equation}
Summing up all the operators, eq.(\ref{eqn:operator1}) becomes
\begin{equation}
  \begin{array}{lcl}
   &&O_1 = \frac{2}{3}\left[
             (\bar{d}_{\alpha}s_{\alpha})_{V-A}
                  (\bar{u}_{\beta}u_{\beta})_{V-A} 
         -(\bar{u}_{\alpha}s_{\alpha})_{V-A}
                   (\bar{d}_{\beta}u_{\beta})_{V-A}\right]
    \\
   &&O_2 = \frac{4}{3}\left[
           (\bar{d}_{\alpha}s_{\alpha})_{V-A}
                 (\bar{u}_{\beta}u_{\beta})_{V-A} 
         +(\bar{u}_{\alpha}s_{\alpha})_{V-A}
               (\bar{d}_{\beta}u_{\beta})_{V-A}\right.
    \\
       &&\qquad\qquad\left.
           +2(\bar{d}_{\alpha}s_{\alpha})_{V-A}
            (\bar{d}_{\beta}d_{\beta})_{V-A}
    +2(\bar{d}_{\alpha}s_{\alpha})_{V-A}
             (\bar{s}_{\beta}s_{\beta})_{V-A}\right]
    \\
   &&O_3 \left( \Delta I=\frac{1}{2} \right) =\frac{4}{9}
      \left[(\bar{d}_{\alpha}s_{\alpha})_{V-A}
            (\bar{u}_{\beta}u_{\beta})_{V-A} 
           +(\bar{u}_{\alpha}s_{\alpha})_{V-A}
            (\bar{d}_{\beta}u_{\beta})_{V-A}\right.
    \\
       &&\qquad\qquad\qquad\qquad\qquad\left.
          +2(\bar{d}_{\alpha}s_{\alpha})_{V-A}
            (\bar{d}_{\beta}d_{\beta})_{V-A}
          -3(\bar{d}_{\alpha}s_{\alpha})_{V-A}
            (\bar{s}_{\beta}s_{\beta})_{V-A}\right]
    \\
   &&O_4 \left( \Delta I=\frac{3}{2}\right)=\frac{20}{9}
      \left[(\bar{d}_{\alpha}s_{\alpha})_{V-A}
            (\bar{u}_{\beta}u_{\beta})_{V-A} 
           +(\bar{u}_{\alpha}s_{\alpha})_{V-A}
            (\bar{d}_{\beta}u_{\beta})_{V-A}
           \right.
    \\
       &&\left.\qquad\qquad\qquad\qquad\qquad
           -(\bar{d}_{\alpha}s_{\alpha})_{V-A}
            (\bar{d}_{\beta}d_{\beta})_{V-A}
           \right]
    \\
   &&O_5 = O_{51}+O_{52}
    \\
   &&\qquad O_{51} = (\bar{d}_{\alpha}s_{\alpha})_{V-A}
           (\bar{u}_{\beta}u_{\beta}+\bar{d}_{\beta}d_{\beta}
            +\bar{s}_{\beta}s_{\beta})_{V+A}
    \\
   &&\qquad O_{52}=-\frac{2}{3}\left[
        (\bar{d}_{\alpha}u_{\alpha})_{S+iP}
        (\bar{u}_{\beta}u_{\beta})_{S-iP}
       +(\bar{d}_{\alpha}d_{\alpha})_{S+iP}
        (\bar{d}_{\beta}s_{\beta})_{S-iP}\right.
    \\
       &&\left.\qquad\qquad\qquad\qquad\qquad
       +(\bar{d}_{\alpha}s_{\alpha})_{S+iP}
        (\bar{s}_{\beta}s_{\beta})_{S-iP}\right]
    \\
   &&O_6=O_{60}+O_{61}
    \\
   &&\qquad O_{60} = (\bar{d}_{\alpha}s_{\beta})_{V-A}
           (\bar{u}_{\beta}u_{\alpha}+\bar{d}_{\beta}d_{\alpha}
            +\bar{s}_{\beta}s_{\alpha})_{V+A}
    \\
   &&\qquad O_{61} = -\frac{32}{9}\left[
           (\bar{d}_{\alpha}u_{\alpha})_{S+iP}
           (\bar{u}_{\beta}s_{\beta})_{S-iP}
         + (\bar{d}_{\alpha}d_{\alpha})_{S+iP}
           (\bar{d}_{\beta}s_{\beta})_{S-iP}\right.
    \\
       &&\left.\qquad\qquad\qquad\qquad\qquad
         + (\bar{d}_{\alpha}s_{\alpha})_{S+iP}
           (\bar{s}_{\beta}s_{\beta})_{S-iP}\right].
  \end{array}
  \label{eqn:operator2}
\end{equation}
The operator $O_{52}$ and $O_{61}$ include $S\pm iP$ currents. The
operator $O_{60}$ have the color non-singlet current. Therefore the
operator $O_{60}$ cannot be replaced by the product of the hadronic
currents.

Next we derive the hadronic currents for the third ansatz. Consider
an extended QCD Lagrangian coupling to external Hermitian matrix
valued fields ${\cal V}_{\mu}$, ${\cal A}_{\mu}$, ${\cal S}_{\mu}$ and 
${\cal P}_{\mu}$;
\begin{equation}
  {\cal L}_{QCD}
   = {\cal L}_{QCD}^0
      +\bar{q}\gamma^{\mu}({\cal V}_{\mu}+\gamma_5 {\cal A}_{\mu})q
    -\bar{q}_R({\cal S}-i{\cal P})q_L -\bar{q}_L({\cal S}+i{\cal P})q_R.
  \label{eqn:qcdlag}
\end{equation}
The chiral Lagrangian and the QCD Lagrangian which can describe the
same hadron phenomena are connected via the external fields. We derive 
the hadron operator currents by taking appropriate derivertives with
respect to the external fields;
\begin{equation}
  \begin{array}{lcl}
 {\cal J}^{ij}_{L\mu}
  =\frac{\partial {\cal L}}{\partial({\cal V}_{\mu}-{\cal A}_{\mu})} &=&
   \bar{q}_i\gamma_{\mu}(1-\gamma_5)q_j 
  \\
  &=&\mbox{Tr}v_{\mu}\bar{B}_v\left[\xi^{\dag}h_{ij}\xi,B_v \right]
  -2D\mbox{Tr}\bar{B}_v S_{v \mu}\left\{\xi^{\dag}h_{ij}\xi,B_v \right\}
   \\
  &&-2F\mbox{Tr}\bar{B}_v S_{v \mu}\left[\xi^{\dag}h_{ij}\xi,B_v\right]
  \\
  &&-v_{\mu}\bar{T}^{\nu}_v\left( \xi^{\dag} h_{ij}\xi \right)T_{v\nu}
  -2{\cal H}\bar{T}^{\nu}_v S_{v\mu} \left( \xi^{\dag}h_{ij}\xi \right)T_{v\nu}
   \\
  &&-2{\cal C}\left(\bar{T}_{v\mu} \left(\xi^{\dag} h_{ij}\xi\right) B_v
     +\bar{B}_v \left(\xi^{\dag} h_{ij}\xi \right) T_{v\mu}\right)
  \\
  &&+if^2_{\pi}\mbox{Tr}
        \left(h_{ij}(\partial_{\mu}\Sigma)\Sigma^{\dag}\right),
  \end{array}
  \label{eqn:left-c}
\end{equation}
\begin{equation}
  \begin{array}{lcl}
 {\cal J}^{ij}_{R\mu}
  =\frac{\partial {\cal L}}{\partial({\cal V}_{\mu}+{\cal A}_{\mu})} &=&
  \bar{q}_i\gamma_{\mu}(1+\gamma_5)q_j
  \\
  &=&\mbox{Tr}v_{\mu}\bar{B}_v\left[\xi h_{ij}\xi^{\dag},B_v \right]
  +2D\mbox{Tr}\bar{B}_v S_{v \mu}\left\{\xi h_{ij}\xi^{\dag},B_v \right\}
   \\
  &&+2F\mbox{Tr}\bar{B}_v S_{v \mu}\left[\xi h_{ij}\xi^{\dag},B_v\right]
  \\
  &&-v_{\mu}\bar{T}^{\nu}_v\left( \xi h_{ij}\xi^{\dag} \right)T_{v\nu}
  +2{\cal H}\bar{T}^{\nu}_v S_{v\mu} 
   \left( \xi h_{ij}\xi^{\dag} \right)T_{v\nu}
   \\
  &&+2{\cal C}\left(\bar{T}_{v\mu} \left(\xi h_{ij}\xi^{\dag}\right) B_v
     +\bar{B}_v \left(\xi h_{ij}\xi^{\dag} \right) T_{v\mu}\right)
  \\
 &&+if^2_{\pi}\mbox{Tr}\left(h_{ij}(\partial_{\mu}\Sigma^{\dag})
    \Sigma\right),
  \end{array}
  \label{eqn:right-c}
\end{equation}
\begin{eqnarray}
 {\cal J}_{(S+iP)}^{ij} =
       {\cal J}_{0(S+iP)}^{ij} &+& {\cal J}_{2(S+iP)}^{ij} =
   \frac{\partial {\cal L}}{\partial ({\cal S}+i{\cal P})} =
   \bar{q}_i(1+\gamma_5 )q_j
  \\
  {\cal J}_{0(S+iP)}^{ij} &=&
  a_1\mbox{Tr}\bar{B}_v\left\{\xi^{\dag}h_{ij}\xi^{\dag},B_v\right\}
  +a_2\mbox{Tr}\bar{B}_v\left[\xi^{\dag}h_{ij}\xi^{\dag},B_v\right]
  \nonumber\\
  &&+a_3\mbox{Tr}\left(\bar{B}_vB_v\right)
   \mbox{Tr}\left(\xi^{\dag}h_{ij}\xi^{\dag}\right)
  \nonumber\\
  &&+c_1\bar{T}^{\mu}_v \left(\xi^{\dag}h_{ij}\xi^{\dag}\right)T_{v\mu}
    +c_2\bar{T}^{\mu}_vT_{v\mu}
   \mbox{Tr}\left(\xi^{\dag}h_{ij}\xi^{\dag}\right)
   \\
  &&+\frac{f^2_{\pi}}{2}B_0\mbox{Tr}\left(h_{ij}\Sigma^{\dag}\right)
   \nonumber\\
  {\cal J}_{2(S+iP)}^{ij} &=& 2B_0L_4
   \mbox{Tr}\left({\cal D}_{\mu}\Sigma{\cal D}^{\mu}\Sigma^{\dag}\right)
    \mbox{Tr}\left(h_{ij}\Sigma^{\dag}\right)
  \nonumber\\
  &&+2B_0L_5\mbox{Tr}\left(h_{ij}
      {\cal D}_{\mu}\Sigma{\cal D}^{\mu}\Sigma^{\dag}\Sigma^{\dag}\right),
  \label{eqn:plus-c}
\end{eqnarray}
\begin{eqnarray}
 {\cal J}_{(S-iP)}^{ij} =
   {\cal J}_{0(S-iP)}^{ij} &+& {\cal J}_{2(S-iP)}^{ij}
  = \frac{\partial {\cal L}}{\partial({\cal S}-i{\cal P})} =
   \bar{q}_i(1-\gamma_5)q_j
  \\
  {\cal J}_{0(S-iP)}^{ij} &=& 
   a_1 \mbox{Tr}\bar{B}_v \bigg\{ \xi h_{ij}\xi,B_v \bigg\}
  +a_2 \mbox{Tr}\bar{B}_v \bigg[\xi h_{ij}\xi,B_v \bigg]
  \nonumber\\
  &&+a_3 \mbox{Tr} \bigg( \bar{B}_vB_v \bigg)
   \mbox{Tr} \bigg( \xi h_{ij}\xi \bigg)
  \nonumber\\
  &&+c_1\bar{T}^{\mu}_v \bigg(\xi h_{ij}\xi\bigg)T_{v\mu}
    +c_2\bar{T}^{\mu}_vT_{v\mu}
   \mbox{Tr}\bigg(\xi h_{ij}\xi\bigg)
  \\
  &&+\frac{f^2_{\pi}}{2}B_0\mbox{Tr}\left(h_{ij}\Sigma\right)
  \nonumber\\
 {\cal J}_{2(S-iP)}^{ij} &=&
   2B_0L_4
   \mbox{Tr}\left({\cal D}_{\mu}\Sigma{\cal D}^{\mu}
           \Sigma^{\dag}\right)\mbox{Tr}\left(h_{ij}\Sigma\right)
  \nonumber\\
  &&+2B_0L_5\mbox{Tr}\left(h_{ij}
   {\cal D}_{\mu}\Sigma{\cal D}^{\mu}\Sigma^{\dag}\Sigma\right),
  \label{eqn:minus-c}
\end{eqnarray}
where $h$ is the $3\times 3$ matrix satisfying the relation
\begin{equation}
  \left(h_{ij}\right)_{ab}=\left\{
  \begin{array}{ll}
  1&a=i\;\mbox{and}\;b=j\\
  0&\mbox{others.}
  \end{array}
  \right.
  \label{eqn:matrix-h}
\end{equation}
These currents are transformed under chiral transform as follows,
\begin{eqnarray}
  {\cal J}_L &:& h_{ij}\longrightarrow Lh_{ij} L^{\dag}
     \qquad (8_L,1_R)
  \nonumber\\
  {\cal J}_R &:& h_{ij}\longrightarrow Rh_{ij} R^{\dag}
     \qquad (1_L,8_R)
  \nonumber\\
  {\cal J}_{(S+iP)} &:& h_{ij}\longrightarrow Lh_{ij} R^{\dag}
     \qquad (3_L,\bar{3}_R)
  \nonumber\\
  {\cal J}_{(S-iP)} &:& h_{ij}\longrightarrow Rh_{ij} L^{\dag}
     \qquad (\bar{3}_L,3_R).
  \nonumber
\end{eqnarray}
Because ${\cal J}^{ij}_{L\mu}$ and ${\cal J}^{ij}_{R\mu}$ are the
Noether currents, which are conserved, they are not renormalized,
while the scalar and pseudo-scalar currents, which are not conserved,
may be renormalized.

Substituting the quark bilinears in the operators
(\ref{eqn:operator2}) by hadronic currents, we obtain a hadronic
representation of the weak Hamiltonian,
\begin{equation}
 {\cal H}^{\Delta S=1}_{eff} =
     -\frac{G_f}{\sqrt{2}}\sum_{r=1}^{6}K_rO_r.
  \label{eqn:weak5}
\end{equation}
The effective weak operators are given by
\begin{equation}
  \begin{array}{lcl}
  O_1 &=& \frac{2}{3}\left({\cal J}^{23}_{L\mu}{\cal J}^{11\mu}_{L}
         -{\cal J}^{13}_{L\mu}{\cal J}^{21\mu}_{L}\right)
  \\
  O_2 &=& \frac{4}{3}\left({\cal J}^{23}_{L\mu}{\cal J}^{11\mu}_{L}
         +{\cal J}^{13}_{L\mu}{\cal J}^{21\mu}_{L}
         +2{\cal J}^{23}_{L\mu}{\cal J}^{22\mu}_{L}
         +2{\cal J}^{23}_{L\mu}{\cal J}^{33\mu}_{L}\right)
  \\
  O_3 &=& O_3\left(\Delta I=1/2 \right)
  \\
      &=& \frac{4}{9}\left({\cal J}^{23}_{L\mu}{\cal J}^{11\mu}_{L}
         +{\cal J}^{13}_{L\mu}{\cal J}^{21\mu}_{L}
         +2{\cal J}^{23}_{L\mu}{\cal J}^{22\mu}_{L}
         -3{\cal J}^{23}_{L\mu}{\cal J}^{33\mu}_{L}\right)
  \\
  O_4 &=& O_3\left(\Delta I=3/2 \right)
  \\
      &=& \frac{20}{9}\left({\cal J}^{23}_{L\mu}{\cal J}^{11\mu}_{L}
         +{\cal J}^{13}_{L\mu}{\cal J}^{21\mu}_{L}
         -{\cal J}^{21}_{L\mu}{\cal J}^{22\mu}_{L}\right)
  \\
  O_5 &=& O_{51}+O_{52}+O_{53}
  \\
  &&O_{51} = \left({\cal J}^{23}_{L\mu}{\cal J}^{11\mu}_{R}
         +{\cal J}^{23}_{L\mu}{\cal J}^{22\mu}_{R}
         +{\cal J}^{23}_{L\mu}{\cal J}^{33\mu}_{R}\right)
  \\
  &&O_{52} = -\frac{2}{3}\left({\cal J}^{21}_{0(S+iP)}{\cal J}^{13}_{0(S-iP)}
         +{\cal J}^{22}_{0(S+iP)}{\cal J}^{23}_{0(S-iP)}
         +{\cal J}^{23}_{0(S+iP)}{\cal J}^{33}_{0(S-iP)}\right)
  \\
  &&O_{53} = -\frac{2}{3}\left({\cal J}^{21}_{2(S+iP)}{\cal J}^{13}_{0(S-iP)}
         +{\cal J}^{21}_{0(S+iP)}{\cal J}^{13}_{2(S-iP)}
         +{\cal J}^{22}_{2(S+iP)}{\cal J}^{23}_{0(S-iP)}\right.
  \\
         &&\left. \qquad\quad
         +{\cal J}^{22}_{0(S+iP)}{\cal J}^{23}_{2(S-iP)}
         +{\cal J}^{23}_{2(S+iP)}{\cal J}^{33}_{0(S-iP)}
         +{\cal J}^{23}_{0(S+iP)}{\cal J}^{33}_{2(S-iP)}\right)
  \nonumber\\
  O_6 &=& O_{61}+O_{62}
  \\
  &&O_{61} = -\frac{32}{9}\left({\cal J}^{21}_{0(S+iP)}{\cal J}^{13}_{0(S-iP)}
         +{\cal J}^{22}_{0(S+iP)}{\cal J}^{23}_{0(S-iP)}
         +{\cal J}^{23}_{0(S+iP)}{\cal J}^{33}_{0(S-iP)}\right)
  \\
  &&O_{62} = -\frac{32}{9}\left({\cal J}^{21}_{2(S+iP)}{\cal J}^{13}_{0(S-iP)}
         +{\cal J}^{21}_{0(S+iP)}{\cal J}^{13}_{2(S-iP)}
         +{\cal J}^{22}_{2(S+iP)}{\cal J}^{23}_{0(S-iP)}\right.
  \\
         && \left. \qquad\quad
         +{\cal J}^{22}_{0(S+iP)}{\cal J}^{23}_{2(S-iP)}
         +{\cal J}^{23}_{2(S+iP)}{\cal J}^{33}_{0(S-iP)}
         +{\cal J}^{23}_{0(S+iP)}{\cal J}^{33}_{2(S-iP)}\right)
  \end{array}
  \label{eqn:operator3}
\end{equation}
The coefficients $K_r$ are given in Table~2.1.

In this study the effective Hamiltonian is constructed within chiral
order ${\cal O}(p^2)$. The quark condensation term has chiral order
${\cal O}(p^0)$. In order to have the consistent chiral ordering to
the effective Hamiltonian, the next-to leading order ${\cal O}(p^2)$
of the scalar and pseudo-scalar currents are needed. Therefore, the
currents ${\cal J}^{ij}_{(S+iP)}$, ${\cal J}^{ij}_{(S-iP)}$ are
divided into ${\cal J}^{ij}_{0(S+iP)}$, ${\cal J}^{ij}_{2(S+iP)}$ and
${\cal J}^{ij}_{0(S-iP)}$, ${\cal J}^{ij}_{2(S-iP)}$,
respectively. The operators $O_5$ and $O_6$ in eq.(\ref{eqn:operator2}) 
are represented by $O_{51}+O_{52}+O_{53}$ and $O_{61}+O_{62}$,
respectively. The constants $a_1$, $a_2$, $a_3$, $c_1$ and $c_2$ are
determined by the comparison between the computation of the chiral
perturbation theory and experimental data.

The above effective weak Hamiltonian includes only the interaction
between color singlet hadron currents which is shown in
Fig.~3.1(a). There is a weak interaction that four quark vertex
appears in the hyperon, which is shown in Fig.~3.1(b). It corresponds
to the two point vertex of the baryon operator in chiral perturbation
theory. It is not enough to extract all the effects of this
interaction by the factorization method from
eq. (\ref{eqn:trans7}). Therefore other terms have to be introduced to
the effective weak Hamiltonian. These terms have the many coupling
constants which are determined by the experimental data. In order to
clear the hadronic currents effect of the weak Hamiltonian, we adopt
the simple effective weak Hamiltonian (\ref{eqn:weak5}). The effect of 
the two point vertex of the baryon operator in the chiral perturbation
theory is discussed in ref.\cite{Takayama,Takayama98}.
It is also noted that
the weak interaction shown in Fig.~3.1(b) do not affect the
$\Delta I=3/2$ amplitudes according to the Pati and Woo
theorem\cite{Pati71}. Hence the $\Delta I=3/2$ amplitudes are appear
only in the operator $O_4$ and we can analyze them. In the following,
we calculate the hyperon non-leptonic weak decay amplitudes with the
effective Hamiltonian (\ref{eqn:weak5}).


\setcounter{equation}{0}
\setcounter{figure}{0}
\section{Hyperon Non-leptonic Weak Decays}
\subsection{Numerical analysis}

Non-leptonic weak decay amplitude is conventionally defined by the
following formula\cite{Data}.
\begin{equation}
  {\cal M}(B_i \longrightarrow B_f+\pi)
   =G_Fm_{{\pi}^+}^2\bar{u}_f(A+B\gamma_5)u_i,
  \label{eqn:ampform1}
\end{equation}
where $A$ is the S-wave amplitude and $B$ is the P-wave one. In the
heavy baryon formalism the decay amplitude is given by
\begin{equation}
  {\cal M}(B_i \longrightarrow B_f+\pi)
   =\frac{G_f}{\sqrt{2}}
     \bar{U}_f({\cal A}+(q\cdot S_v){\cal B})U_i,
  \label{eqn:ampform2}
\end{equation}
where $q$ is the momentum of outgoing pion. Note that $G_F$ and $G_f$
are the Fermi coupling constants but have different values:
$G_F=10^{-5}m_p^{-2}$, $G_f=1.166\times 10^{-11}[\mbox{MeV}^{-2}]$.
In the heavy baryon formalism, the baryons have the two component
spinors, while baryons in eq. (\ref{eqn:ampform1}) are described by
four component spinors. In the rest frame of the initial baryon, the
two component spinors are extracted from eq. (\ref{eqn:ampform1}), and
we obtain the following relations between decay amplitudes, 
\begin{equation}
  \begin{array}{lcl}
   A &=& \frac{-iG_f}{\sqrt{2}G_Fm_{{\pi}^+}^2}f_{\pi}{\cal A}
  \\
   B &=& \frac{iG_f}{2\sqrt{2}G_Fm_{{\pi}^+}^2}
         (E_f+m_f)f_{\pi}{\cal B},
  \end{array}
  \label{eqn:ampform3}
\end{equation}
where $E_f$ is the energy of the final baryon and $m_f$ is the final
state baryon mass. The hyperon non-leptonic weak decays are measured
for the following seven processes:
\\
$\Sigma^- \longrightarrow n+\pi^-\;\;\left(\Sigma^-_-\right),\;\;\;
 \Sigma^+ \longrightarrow n+\pi^+\;\;\left(\Sigma^+_+\right),\;\;\;
 \Sigma^+ \longrightarrow p+\pi^0\;\;\left(\Sigma^+_0\right),$
\\
$\Lambda \longrightarrow n+\pi^0\;\;\left(\Lambda^0_0\right),\;\;\;
 \Lambda \longrightarrow p+\pi^-\;\;\left(\Lambda^0_-\right),\;\;\;
 \Xi^- \longrightarrow \Lambda+\pi^-\;\;\left(\Xi^-_-\right),\;\;\;
 \Xi^0 \longrightarrow \Lambda+\pi^0\;\;\left(\Xi^0_0\right)$.
\\
There are 14 measured amplitudes since each 7 decay process have the
S- and P-wave amplitudes.

Using the strong interaction Lagrangian (\ref{eqn:mesonlag1}),
(\ref{eqn:mesonlag2}), (\ref{eqn:baryonlag1}) and the effective weak
Hamiltonian (\ref{eqn:weak5}), the S- and P-wave hyperon non-leptonic
decay amplitudes are calculated. The tree level amplitudes are
obtained from the calculation of Feynman diagrams in Fig.~4.1. Since
the diagrams (2), (3) and (4) have a strong interaction vertex
$(q\cdot S_v)$, these diagrams cause only P-wave decay amplitudes. The 
chiral logarithmic terms in the one-loop correction are obtained from
the calculation of Feynman diagrams in Fig.~4.2. The S-wave amplitudes 
are obtained from Fig.~4.2 (9), (10), (11), (30), (35), (40), (41),
(42), (43), (44), (45), (48). The rest of diagrams and (40), (41),
(42), (43), (44), (45) contribute to the chiral logarithmic terms of
the P-wave amplitudes. The wave function renormalization is applied to 
the internal and external lines of hadrons. Examples of the wave
function renormalization are shown in Fig.~4.3. The amplitudes are
given by the summation of these diagrams. In our study the wave
function renormalization for the intermediate baryon and meson are
taken into account. The tree level amplitudes are given by
\\
%
S-wave:
\begin{equation}
  \begin{array}{lcl}
A(\Sigma^-_-) &=& 
  3.0705\times 10^{-1} - 2.3337\times 10^{-1}a_1 + 2.3337\times 10^{-1}a_2
  \\
A(\Sigma^+_+) &=& 0
  \\
A(\Sigma^+_0) &=& 
  -7.0563\times 10^{-2} + 1.6502\times 10^{-1}a_1 - 1.6502\times 10^{-1}a_2
  \\
A(\Lambda^0_0) &=& 
  -6.3084\times 10^{-2} - 6.7368\times 10^{-2}a_1 - 2.0210\times 10^{-1}a_2
  \\
A(\Lambda^-_-) &=& 
  2.7450\times 10^{-1} + 9.5273\times 10^{-2}a_1 + 2.8582\times 10^{-1}a_2
  \\
A(\Xi^-_-) &=& 
  -3.1012\times 10^{-1} + 9.5273\times 10^{-2}a_1 - 2.8582\times 10^{-1}a_2
  \\
A(\Xi^0_0) &=& 
  7.1268\times 10^{-2} - 6.7368\times 10^{-2}a_1 + 2.0210\times 10^{-1}a_2
  \end{array}
    \label{eqn:stree}
\end{equation}
P-wave:
\begin{equation}
  \begin{array}{lcl} 
B(\Sigma^-_-) &=& 
  2.2212 + 1.2058a_1 + 8.2856\times 10^{-1}a_2
  \\
B(\Sigma^+_+) &=& -1.9490a_1 + 3.9833a_2
  \\
B(\Sigma^+_0) &=&
  -1.2804 - 2.2308a_1 + 2.2308a_2
  \\ 
B(\Lambda^0_0) &=&
  4.4732 - 1.1459\times 10^{-1}a_1 - 2.7816a_2
  \\ 
B(\Lambda^0_-) &=&
  -7.7599 + 1.6205\times 10^{-1}a_1 + 3.9338a_2
  \\
B(\Xi^-_-) &=&
  3.0086 - 1.8612a_1 - 2.7488a_2
  \\
B(\Xi^0_0) &=&
  -1.7343 + 1.3161a_1 + 1.9437a_2
  \end{array}
  \label{eqn:ptree}
\end{equation}
The tree level amplitudes and the chiral logarithmic terms in the
one-loop correction are given by
\\
%
S-wave:
\begin{equation}
  \begin{array}{lcl}
A(\Sigma^-_-) &=& 
  3.2554\times 10^{1} + 3.0862\times 10^{1}a_1
  - 3.1032\times 10^{1}a_2 - 2.5229\times 10^{-2}c_1
  \\
A(\Sigma^+_+) &=&
  -6.0803 - 1.0944\times 10^{-4}a_1 + 6.7828\times 10^{-3}a_2
  \\
A(\Sigma^+_0) &=&
  -1.5360\times 10^{1} - 2.1816\times 10^{1}a_1 + 2.1951\times 10^{1}a_2 + 
   1.7840\times 10^{-2}c_1
  \\
A(\Lambda^0_0) &=&
  -3.1614\times 10^{-1} + 8.8775a_1 + 2.6754\times 10^{1}a_2 + 
   1.0109\times 10^{-1}c_1
  \\
A(\Lambda^0_-) &=&
  2.5155 - 1.2557\times 10^{1}a_1 - 3.7855\times 10^{1}a_2 - 
   1.4297\times 10^{-1}c_1
  \\
A(\Xi^-_-) &=&
  -2.4018\times 10^{1} - 1.2758\times 10^{1}a_1 + 3.7914\times 10^{1}a_2 + 
   5.3009\times 10^{-2}c_1
  \\
A(\Xi^0_0) &=&
  1.1550\times 10^{1} + 9.0225a_1 - 2.6800\times 10^{1}a_2 -
   3.7483\times 10^{-2}c_1
  \end{array}
    \label{eqn:sloop}
\end{equation}
P-wave:
{\small
\begin{equation}
  \begin{array}{lcl} 
B(\Sigma^-_-) &=& 
  6.9622\times 10^{2} + 2.2896\times 10^{1}a_1 + 2.3957\times 10^{1}a_2 +
   8.3649a_3 - 5.4327\times 10^{-1}c_1
  \\
B(\Sigma^+_+) &=&
  1.4794\times 10^{2} - 3.2295\times 10^{2}a_1 + 2.8974\times 10^{2}a_2 - 
   7.2823\times 10^{-1}c_1
  \\
B(\Sigma^+_0) &=&
  -3.8548\times 10^{2} - 2.4455\times 10^{2}a_1 + 1.8794\times 10^{2}a_2 -
   5.9149a_3 - 1.3079\times 10^{-1}c_1
  \\
B(\Lambda^0_0) &=&
  1.7307\times 10^{3} - 7.6621\times 10^{1}a_1 - 3.2130\times 10^{2}a_2 +
   1.6514\times 10^{1}a_3 - 5.3588\times 10^{-1}c_1
  \\
B(\Lambda^0_-) &=&
  -2.4466\times 10^{3} + 1.0836\times 10^{2}a_1 + 4.5438\times 10^{2}a_2 -
   2.3354\times 10^{1}a_3 + 7.5786\times 10^{-1}c_1
  \\
B(\Xi^-_-) &=&
  9.6124\times 10^{2} - 2.9146\times 10^{2}a_1 - 3.0700\times 10^{2}a_2 +
  9.9536a_3 - 3.8293\times 10^{-1}c_1
  \\
B(\Xi^0_0) &=&
  -6.8022\times 10^{2} + 2.0609\times 10^{2}a_1 + 2.1708\times 10^{2}a_2 -
   7.0383a_3 + 2.7077\times 10^{-1}c_1
  \end{array}
  \label{eqn:ploop}
\end{equation}
}
In this computation, following hadron masses and constants are used:
\begin{equation}
  \begin{array}{ccccccccc}
  m_n &=& 939\mbox{[MeV]},&m_{\Sigma} &=& 1193\mbox{[MeV]},
  &m_{\Lambda} &=& 1116\mbox{[MeV]}\\
  m_{\Xi} &=& 1318\mbox{[MeV]},&m_{\Delta} &=& 1232\mbox{[MeV]},
  &m_{\Sigma^*} &=& 1385\mbox{[MeV]}\\
  m_{\Xi^*} &=& 1533\mbox{[MeV]}, &m_{\Omega} &=& 1672\mbox{[MeV]}
  &&&\\
  M_{\pi} &=& 138\mbox{[MeV]},& M_K &=& 496\mbox{[MeV]},
  &M_{\eta} &=& 547\mbox{[MeV]}\\
  D &=& 0.61,&F &=& 0.40,&{\cal C} &=& 1.6\\
  {\cal H} &=& -1.9, &f_{\pi} &=& 93\mbox{[Mev]}, 
  &L_4&=&-0.3003\\
  L_5&=&1.399,&B_0&=&1391.4\mbox{[MeV]},&&&\\
  \mbox{data set 2.}&&&&&&&& \\
  \end{array}
  \label{eqn:constants1}
\end{equation}
And for the chiral logarithmic terms, the renormalization scale
$4\pi\mu^2=1.0\times 10^6 [\mbox{Mev}^2]$ is used.
The amplitudes have 4 unknown parameters, $a_1$, $a_2$, $a_3$ and
$c_1$. The constants $a_1$ and $a_2$ that appear in the scalar
current are determined as $a_1=29.1/m_s$ and $a_2=-94.8/m_s$ where
$m_s$ is the strange quark mass. They are derived from the baryon mass 
difference with the chiral perturbation theory\cite{Jenkins92a}. The
remaining two parameters are fitted with 14 experimental data. The
amplitudes derived from fitted parameters are shown in Table~4.1. The
fitted parameters are shown in Table~4.2.
Constant terms in eq. (\ref{eqn:sloop}) and (\ref{eqn:ploop}) have
large values, which are caused by the quark condensation value
$B_0$. The decay amplitudes depend on the quark condensation value
strongly. To reproduce the experimental data, the quark
condensation value is changed as $B_0=1.3914\times 10^3\mbox{[MeV]}$,
$1.3914\times 10^2\mbox{[MeV]}$, $0.0\mbox{[MeV]}$. Since the Table~4.1 
shows that amplitudes extracted with data set 2 and
$B_0=1.3914\times 10^2\mbox{[MeV]}$ 
have better fitting with experimental data, we adopt these parameters
in the following analysis.
From the Table~2.1, the difference between data set 1 and 2 are large
in $K_5$ and $K_6$. This difference is responsible for the amplitudes
caused by the operators $O_5$ and $O_6$.  
The components of the predicted amplitudes are shown in Table~4.3.
The operators $O_{52}$, $O_{53}$, $O_{61}$ and $O_{62}$ which include
the scalar and pseudo-scalar currents are proportional to the quark
condensation value $B_0$. If we adopt the condition
$B_0=0.0\mbox{[MeV]}$, The amplitudes do not depend on the parameters
$a_1$, $a_2$, $a_3$ and $c_1$. The components of these amplitudes are 
shown in Table~4.4.
These tables suggest that the contributions of the chiral logarithmic
correction are large. The tree level amplitudes caused by each
operator in eq. (\ref{eqn:operator3}) are shown in Table~4.5 and the
chiral logarithmic amplitudes caused by each operator are shown in
Table~4.6.


\subsection{Discussion}

Table~4.1 shows that the amplitudes derived by the hadronic current
interaction do not reproduce the experimental data well. It is caused
by the lack of two point vertex of the baryon operator and by
uncertainties of the quark condensation in scalar and pseudo-scalar
currents. If we choose the condition $B_0=0.0\mbox{[MeV]}$, the
contributions of the scalar and pseudo-scalar currents become zero.
The amplitudes are, therefore, caused by the $V\pm A$ current
interaction in the operator $O_1$, $O_2$, $O_3$, $O_4$ and $O_{51}$.
Table~4.1 shows that the S-wave amplitudes $\Sigma^-_-$, 
$\Lambda^0_-$ and $\Xi^-_-$ are almost caused by the $V\pm A$ currents.
In the other amplitudes the contributions of the scalar, pseudo-scalar
currents and two point vertex of the baryon operator are large.

Table~4.3 and 4.4 show the large amplitudes caused by the chiral
logarithmic correction.
In the chiral order analysis, the logarithmic correction of the one-loop 
diagram has the order of magnitude,
$\frac{M^2_K}{16\pi^2f_{\pi}^2}\ln\left(\frac{M^2_K}{4\pi\mu^2}\right)
\approx 0.20$
The effect of the one-loop calculation expected to be $25\%$ correction
to tree level amplitudes.
But there are many loop diagrams which have many internal baryon states.
Since the amplitudes are obtained by the summation of all
diagrams, the amplitudes caused by the loop correction become large.

Let us considering the contribution of the each operator. In the tree
level the operator $O_1$ is most important for the amplitudes, which
is shown in Table~4.5. Since the tree level amplitudes do not depend on
the free parameters $a_3$ and $c_1$, these amplitudes depend on $B_0$
value and they are small. The amplitudes caused by the two point vertex
of the baryon operator must be large. But in the chiral logarithmic
correction, the operators composed of the scalar and pseudo-scalar
currents become important. Especially the weak Hamiltonian caused by
meson currents, which is shown in Fig.\ref{fig:loop} (9), (10) etc., is
important since these terms are proportional to $B^2_0$ and sensitive to 
the $B_0$ value. In order to reproduce the experimental data by the
chiral perturbation theory, it is important to construct the $O_5$ and
$O_6$ operators exactly. 

In our method $\Delta I=3/2$ amplitudes are derived from the operator
$O_4$ and it does not depend on the unknown coupling constant in the
chiral perturbation theory. 
In the tree level, the S-wave $\Delta I=3/2$ amplitudes are given by the 
diagram (1) in Fig.\ref{fig:tree} and the P-wave amplitudes are given by 
the diagrams (1) and (4) in Fig.\ref{fig:tree}. The chiral logarithmic
corrections of the S-wave $\Delta I=3/2$ amplitudes are obtained from
the diagrams (9), (10), (11), (30), (35), (40), (41) and (48) in
Fig.\ref{fig:loop}, and the P-wave amplitudes are obtained from the
diagrams (1)$\sim$(8), (26)$\sim$(29), (31)$\sim$(34),
(36)$\sim$(41) and (48) in Fig.\ref{fig:loop}.
The comparison between the experimental
data and the calculated data with our method are shown in
Table~4.7. In the tree level, the absolute values of the $\Delta I=3/2$
amplitudes become small. It is consistent to the $\Delta I=1/2$
rule. These small amplitudes are caused by the small coupling constants
$F,\;D <1$ in the chiral perturbation theory. But including the chiral
logarithmic correction of the one-loop diagrams, the amplitudes become
large. These amplitudes are caused by the summation of the many
logarithmic term in loop diagrams and by the large coupling constants
$|{\cal H}|,\;{\cal C}>1$ of the decuplet fields. 


\setcounter{equation}{0}
\setcounter{figure}{0}
\section{Conclusion}

Hyperon non-leptonic weak decay amplitudes are studied by the chiral
perturbation theory. Applying the renormalization method to the weak
interaction, the effective weak Hamiltonian which has perturbative QCD 
correction is obtained. The effective weak Hamiltonian is described
by the qurak bilinear form. These quark currents are substituted by
the hadronic currents derived by the chiral perturbation theory.

Our results suggest that the color singlet $V\pm A$ current
interaction has the small contribution to the decay amplitudes. The
weak interaction caused by the scalar, pseudo-scalar currents and by
the two point vertex of the baryon operator have the large
contribution. The $\Delta I=3/2$ amplitudes are suppressed. Using the
chiral perturbation theory as the strong interaction correction for the
weak interaction, it is consistent to the experimental data.
Our method have no parameters for the $\Delta I=3/2$ amplitudes.
Applying our method to the more complicated decay process, like
$Y+N\longrightarrow N+N$, it is possible to predict the characteristic
of the $\Delta I=3/2$ amplitudes.

 In order to reproduce the experimental data quite well, it is
necessary to study the two point vertex of the baryon  operator,
the quark condensation value $B_0$ for the weak interaction and the
relation between the renormalization scale $\mu$ of the perturbative QCD 
correction and the chiral perturbation theory.

{\bf\large Table~2.1} \hspace{3mm}
The values of the coefficients in the effective
weak Hamiltonian (\ref{eqn:weak4}). 
The values are taken from Ref. \cite{Paschos90}.
 The data
set 1 corresponds to the choices $m_t=200$[GeV] and $\mu_0=0.24$[GeV],
$\Lambda_{QCD}=0.10$.
The data set 2 corresponds to the choices $m_t=200$[GeV] and
$\mu_0=0.71$[GeV], $\Lambda_{QCD}=0.316$. In both cases, $\mu_0$ is
defined so as to satisfy $\alpha_s(\mu^2)=1$.
\begin{center}
  \begin{tabular}{|c|c|c|}\hline
  &data set 1& data set 2 \\ \hline
  $\mu$(GeV)&0.24&0.71 \\ \hline
  $\Lambda^{(4)}$(GeV)&0.10&0.316 \\ \hline
  \hline
  $K_1$&-0.284&-0.270 \\ \hline
  $K_2$&0.009&0.011 \\ \hline
  $K_3$&0.026&0.027 \\ \hline
  $K_4$&0.026&0.027 \\ \hline
  $K_5$&0.004&0.002 \\ \hline
  $K_6$&0.004&0.002 \\ \hline
  \end{tabular}
\end{center}

\vspace{1cm}
\noindent
{\bf\large Table~3.1} \hspace{3mm}
Phenomenological values of the renormalized couplings
$L_i^r(M_{\rho})$.
$L^r_i(i=4,\cdots,8)$ are from ref. \cite{Gasser84},
$L^r_{9,10}$ from ref. \cite{Bijnens88} and
$L^r_{1,2,3}$ from ref. \cite{Bijnens92}.

\begin{center}
\begin{tabular}{|c|c|}\hline
  i&$L_i^r(M_{\rho})$\\ \hline
  1&0.7$\pm$0.5\\
  2&1.2$\pm$0.4\\
  3&-3.6$\pm$1.3\\
  4&-0.3$\pm$0.5\\
  5&1.4$\pm$0.5\\
  6&-0.2$\pm$0.3\\
  7&-0.4$\pm$0.15\\
  8&0.9$\pm$0.3\\
  9&6.9$\pm$0.2\\
  10&-5.2$\pm$0.3\\ \hline
\end{tabular}
\end{center}

\newpage
\noindent{\bf\large Table~4.1} \hspace{3mm}
The tree and one-loop level amplitudes. First column
corresponds to the experimental data. The quark condensation
value $B_0$ is varied as $B_0=1.3914\times 10^3\mbox{[MeV]}$,
$B_0=1.3914\times 10^2\mbox{[MeV]}$ and
$B_0=0.0\mbox{[MeV]}$.\\
\vspace{2mm}\\
{\scriptsize
\begin{center}
\begin{tabular}{|c|c|c|c|c|c|c|c|}
\multicolumn{8}{l}{{\normalsize S-wave:}}\\ \hline
decay&&\multicolumn{3}{|c|}{data set 1}
&\multicolumn{3}{|c|}{data set 2}\\ \cline{3-8}
mode &exp.
&{\scriptsize $1.3941\times 10^3$}
&{\scriptsize $1.3941\times 10^2$}
&{\scriptsize $0.0$}
&{\scriptsize $1.3941\times 10^3$}
&{\scriptsize $1.3941\times 10^2$}
&{\scriptsize $0.0$}\\ \hline
\hline

$\Sigma^-_-$
&$1.93\pm 0.01$
& 93.06
& 6.157
& 1.732
& 45.32
& 4.016
& 1.717\\ \hline

$\Sigma^+_+$
&$0.06\pm 0.01$
& -12.75
& -0.1616
& -0.03402
& -6.082
& -0.09894
& -0.03831\\ \hline

$\Sigma^+_0$
&$-1.48\pm 0.05$
& -50.80
& -3.278
& -0.2901
& -24.39
& -1.805
& -0.2463\\ \hline

$\Lambda^0_0$
&$-1.07\pm 0.02$
& -14.24
& -6.484
& -0.5509
& -7.790
& -4.1018
& -0.4985\\ \hline

$\Lambda^0_-$
&$1.47\pm 0.01$
& 23.13
& 10.35
& 1.938
& 13.09
& 7.012
& 1.907\\ \hline

$\Xi^-_-$
&$-2.04\pm 0.01$
& -75.01
& -7.689
& -2.173
& -37.17
& -5.175
& -2.142\\ \hline

$\Xi^0_0$
&$1.54\pm 0.03$
& 42.63
& 4.480
& 0.6749
& 20.85
& 2.719
& 0.6203\\ \hline
\end{tabular}
\end{center}
}
{\footnotesize
\begin{center}
\begin{tabular}{|c|c|c|c|c|c|c|c|}
\multicolumn{8}{l}{{\normalsize P-wave:}}\\ \hline
decay&&\multicolumn{3}{|c|}{data set 1}
&\multicolumn{3}{|c|}{data set 2}\\ \cline{3-8}
mode &exp.
&{\scriptsize $1.3941\times 10^3$}
&{\scriptsize $1.3941\times 10^2$}
&{\scriptsize $0.0$}
&{\scriptsize $1.3941\times 10^3$}
&{\scriptsize $1.3941\times 10^2$}
&{\scriptsize $0.0$}\\ \hline
\hline

$\Sigma^-_-$
&$-0.65\pm 0.07$
& -433.8
& -3.070
& 2.304
& -204.2
& 0.5967
& 2.316\\ \hline

$\Sigma^+_+$
&$19.07\pm 0.07$
& 31.33
& 15.84
& 7.479
& 24.19
& 16.83
& 7.216\\ \hline

$\Sigma^+_0$
&$12.04\pm 0.58$
& 330.8
& 15.75
& 6.038
& 163.7
& 13.95
& 5.935\\ \hline

$\Lambda^0_0$
&$-7.14\pm 0.56$
& 128.9
& -4.52
& -0.5050
& 58.51
& -4.901
& -0.4810\\ \hline

$\Lambda^0_-$
&$9.98\pm 0.24$
& -179.9
& 6.079
& 0.3647
& -81.83
& 6.581
& 0.3174\\ \hline

$\Xi^-_-$
&$6.93\pm 0.31$
& -71.30
& 1.399
& 0.4604
& -34.72
& -0.1677
& 0.4341\\ \hline

$\Xi^0_0$
&$-6.43\pm 0.66$
& 49.51
& -1.167
& -0.4958
& 24.02
& -0.06164
& -0.4837\\ \hline
\end{tabular}
\end{center}
}

\newpage
\vspace{8mm}
\noindent
{\bf\large Table~4.2} \hspace{3mm}
The parameters of scalar and pseudo-scalar currents.
Two types of data set are used. Fixing the
$B_0$ value, we obtain eq. (\ref{eqn:stree}), (\ref{eqn:ptree}),
(\ref{eqn:sloop}) and (\ref{eqn:ploop}).
The parameters are fitted with least square method.

\vspace{3mm}
\noindent
{\small
\begin{center}
\begin{tabular}{|c|c|c|c|} \hline
parameter&\multicolumn{3}{|c|}{dataset 1}\\ \hline
$B_0$&$1.3914\times 10^{3}$&$1.3914\times 10^{2}$&0.0\\ \hline
\hline
$a_1$&
$1.9400\times 10^{-1}$&
$1.9400\times 10^{-1}$&
-\\ \hline
$a_2$&
$-6.3200\times 10^{-1}$&
$-6.3200\times 10^{-1}$&
-\\ \hline
$a_3$&
$-1.0652\times 10^{2}$&
$-2.3960\times 10^{1}$&
-\\ \hline
$c_1$&
$8.9337$&
$-2.0402\times 10^{2}$&
-\\ \hline
\end{tabular}
\end{center}
\begin{center}
\begin{tabular}{|c|c|c|c|} \hline
parameter&\multicolumn{3}{|c|}{dataset 2}\\ \hline
$B_0$&$1.3914\times 10^{3}$&$1.3914\times 10^{2}$&0.0\\ \hline
\hline
$a_1$&
$1.9400\times 10^{-1}$&
$1.9400\times 10^{-1}$&
-\\ \hline
$a_2$&
$-6.3200\times 10^{-1}$&
$-6.3200\times 10^{-1}$&
-\\ \hline
$a_3$&
$-1.0692\times 10^{2}$&
$-2.7993\times 10^{1}$&
-\\ \hline
$c_1$&
$1.1855$&
$-2.8150\times 10^{2}$&
-\\ \hline
\end{tabular}
\end{center}
}

\newpage
\vspace{8mm}
\noindent
{\bf\large Table~4.3}\hspace{3mm}
The predicted amplitudes with
$B_0=1.3941\times 10^2\mbox{[MeV]}$, data set 2 and the parameters in
Table 4.2. $A_{exp}$  and $B_{exp}$ correspond to the experimental
data. $A_{tree}$ and $B_{tree}$ are the tree level amplitudes.
$\Delta A_{loop}$ and $\Delta B_{loop}$ are the summation of the chiral
logarithmic correction of the one-loop graphs. They are represented
with $\Delta A_{loop}=\Delta A_{octet}+\Delta A_{decuplet}$ and
$\Delta B_{loop}=\Delta B_{octet}+\Delta B_{decuplet}$.
$A_{theory}$ and $B_{theory}$ are the total amplitude of the chiral
perturbation theory, which are represented with
$A_{theory}=A_{tree}+\Delta A_{loop}$ and
$B_{theory}=B_{tree}+\Delta B_{loop}$. 
{\footnotesize
\begin{center}
\begin{tabular}{|c|c|c|c|c|c|c|}
\multicolumn{7}{l}{{\normalsize S-wave:}}\\ \hline
decay mode&$A_{exp.}$&$A_{theory}$&$A_{tree}$&$\Delta A_{loop}$
&$\Delta A_{octet}$&$\Delta A_{decuplet}$\\ \hline
$\Sigma_-^-$ &$1.93\pm 0.01$
             &4.016   &0.2974  &3.719   &1.739&1.980 \\ \hline
$\Sigma_+^+$ &$0.06\pm 0.01$ 
             &-0.09894&0.0     &-0.09894&-0.04420&-0.05474 \\ \hline 
$\Sigma^+_0$ &$-1.48\pm 0.05$
             &-1.805  &-0.06375&-1.741  &-1.149&-0.5920 \\ \hline
$\Lambda_0^0$&$-1.07\pm 0.02$
             &-4.102  &-0.05735&-4.044  &-0.6702&-3.374 \\ \hline
$\Lambda^0_-$&$1.47\pm 0.01$
             &7.012   &0.2664  &6.746   &1.619&5.127 \\ \hline
$\Xi^-_-$    &$-2.04\pm 0.01$
             &-5.175  &-0.3002 &-4.875  &-1.880&-2.995 \\ \hline
$\Xi^-_0$    &$1.54\pm 0.03$
             &2.719   &0.06423 &2.655   &1.109&1.546 \\ \hline
\end{tabular}
\end{center}
}
{\footnotesize
\begin{center}
\begin{tabular}{|c|c|c|c|c|c|c|}
\multicolumn{7}{l}{{\normalsize P-wave:}}\\ \hline
decay mode&$B_{exp.}$&$B_{theory}$&$B_{tree}$&$\Delta B_{loop}$
&$\Delta B_{octet}$&$\Delta B_{decuplet}$\\ \hline
$\Sigma_-^-$ &$-0.65\pm 0.07$
             &0.5967 &0.6097  &-0.01300&-17.31&17.29  \\ \hline
$\Sigma_+^+$ &$19.07\pm 0.07$
             &16.83  &-0.1448 &16.98   &-11.31&28.29  \\ \hline 
$\Sigma^+_0$ &$12.04\pm 0.58$
             &13.95  &-0.2433 &14.19   &4.578 &9.615  \\ \hline
$\Lambda_0^0$&$-7.14\pm 0.56$
             &-4.901 &0.6148  &-5.516  &-19.30&13.79  \\ \hline
$\Lambda^0_-$&$9.98\pm 0.24$
             &6.581  &-2.303  &8.885   &26.22 &-17.34 \\ \hline
$\Xi^-_-$    &$6.93\pm 0.31$
             &-0.1677&0.9142  &-1.081  &-10.93&9.851  \\ \hline
$\Xi^-_0$    &$-6.43\pm 0.66$
             &-0.06164&-0.2534&0.1917  &8.392 &-8.200 \\ \hline
\end{tabular}
\end{center}
}

\newpage
\vspace{8mm}
\noindent
{\bf\large Table~4.4}\hspace{3mm}
The predicted amplitudes with
$B_0=0.0\mbox{[MeV]}$, data set 2.
$A_{exp}$  and $B_{exp}$ correspond to the experimental
data. $A_{tree}$ and $B_{tree}$ are the tree level amplitudes.
$\Delta A_{loop}$ and $\Delta B_{loop}$ are the summation of the chiral
logarithmic correction of the one-loop graphs. They are represented
with $\Delta A_{loop}=\Delta A_{octet}+\Delta A_{decuplet}$ and
$\Delta B_{loop}=\Delta B_{octet}+\Delta B_{decuplet}$.
$A_{theory}$ and $B_{theory}$ are the total amplitude of the chiral
perturbation theory, which are represented with
$A_{theory}=A_{tree}+\Delta A_{loop}$ and
$B_{theory}=B_{tree}+\Delta B_{loop}$. 
{\footnotesize
\begin{center}
\begin{tabular}{|c|c|c|c|c|c|c|}
\multicolumn{7}{l}{{\normalsize S-wave:}}\\ \hline
decay mode&$A_{exp.}$&$A_{theory}$&$A_{tree}$&$\Delta A_{loop}$
&$\Delta A_{octet}$&$\Delta A_{decuplet}$\\ \hline
$\Sigma_-^-$ &$1.93\pm 0.01$
             &1.717   &0.3071  &1.410   &0.3558&1.054  \\ \hline
$\Sigma_+^+$ &$0.06\pm 0.01$ 
             &-0.03831&0.0     &-0.03831&-0.04977&0.01146 \\ \hline 
$\Sigma^+_0$ &$-1.48\pm 0.05$
             &-0.2463  &-0.07056&-0.1757  &-0.2162&0.04050 \\ \hline
$\Lambda_0^0$&$-1.07\pm 0.02$
             &-0.4985  &-0.06308&-0.4354  &0.07263&-0.5080 \\ \hline
$\Lambda^0_-$&$1.47\pm 0.01$
             &1.907   &0.2745  &1.633   &0.5580&1.075 \\ \hline
$\Xi^-_-$    &$-2.04\pm 0.01$
             &-2.142  &-0.3101 &-1.832  &-0.4793&-1.353 \\ \hline
$\Xi^-_0$    &$1.54\pm 0.03$
             &0.6203   &0.07127 &0.5490   &0.1440&0.4050 \\ \hline
\end{tabular}
\end{center}
}
{\footnotesize
\begin{center}
\begin{tabular}{|c|c|c|c|c|c|c|}
\multicolumn{7}{l}{{\normalsize P-wave:}}\\ \hline
decay mode&$B_{exp.}$&$B_{theory}$&$B_{tree}$&$\Delta B_{loop}$
&$\Delta B_{octet}$&$\Delta B_{decuplet}$\\ \hline
$\Sigma_-^-$ &$-0.65\pm 0.07$
             &2.316 &0.6080  &1.708&-0.6232&2.331  \\ \hline
$\Sigma_+^+$ &$19.07\pm 0.07$
             &7.216  &0.0 &7.216   &-0.02100&7.237  \\ \hline 
$\Sigma^+_0$ &$12.04\pm 0.58$
             &5.935  &-0.1397 &6.075   &0.7635 &5.312  \\ \hline
$\Lambda_0^0$&$-7.14\pm 0.56$
             &-0.4810 &0.4882  &-0.9692  &0.06545&-1.035  \\ \hline
$\Lambda^0_-$&$9.98\pm 0.24$
             &0.3174  &-2.124  &2.442   &-1.179 &3.621 \\ \hline
$\Xi^-_-$    &$6.93\pm 0.31$
             &0.4341&0.8236  &-0.3894  &0.6008&-0.9902  \\ \hline
$\Xi^-_0$    &$-6.43\pm 0.66$
             &-0.4837&-0.1893&-0.2945  &0.2395 &-0.5340 \\ \hline
\end{tabular}
\end{center}
}

\newpage
\vspace{8mm}
\noindent
{\bf\large Table~4.5}\hspace{3mm}
The amplitudes for each operator in the tree level
calculation with data set 2 and $B_0=1.3914\times10^2\mbox{[MeV]}$.
The first column corresponds to the operator number of the
effective Hamiltonian (\ref{eqn:weak5}) and
(\ref{eqn:operator3}). The second row shows the
observed data.\\
{\footnotesize
\begin{center}
\begin{tabular}{|c|c|c|c|c|c|c|c|}
\multicolumn{8}{l}{{\normalsize S-wave:}}\\ \hline
vertex&$\Sigma_-^-$&$\Sigma_+^+$&$\Sigma^+_0$&$\Lambda_0^0$&
$\Lambda^0_-$&$\Xi^-_-$&$\Xi^-_0$\\ \hline
exp.&1.93
&0.06
&-1.48
&-1.07
&1.47
&-2.04
&-1.54 \\ \hline 
$O_1$
&0.2073
&0
&-0.1466
&-0.1310
&0.1853
&-0.2093
&0.1480
\\ \hline
$O_2$
&0.01689
&0
&-0.01194
&-0.01068
&0.01510
&-0.01706
&0.01206
\\ \hline
$O_3$
&0.01382
&0
&-0.009770
&-0.008735
&0.01235
&-0.01396
&0.009868
\\ \hline
$O_4$
&0.06909
&0
&0.09770
&0.08735
&0.06176
&-0.06978
&-0.09868
\\ \hline
$O_{51}$
&0
&0
&0
&0
&0
&0
&0
\\ \hline
$O_{52}$
&3.755E-4
&0
&-2.655E-4
&-2.234E-4
&3.159E-4
&-3.879E-4
&2.743E-4
\\ \hline
$O_{53}$
&0
&0
&0
&0
&0
&0
&0
\\ \hline
$O_{61}$
&-0.01001
&0
&0.007081
&0.005956
&-0.008424
&0.01034
&-0.007314
\\ \hline
$O_{62}$
&0
&0
&0
&0
&0
&0
&0
\\ \hline
\end{tabular}
\end{center}
}
{\footnotesize
\begin{center}
\begin{tabular}{|c|c|c|c|c|c|c|c|}
\multicolumn{8}{l}{{\normalsize P-wave:}}\\ \hline
vertex&$\Sigma_-^-$&$\Sigma_+^+$&$\Sigma^+_0$&$\Lambda_0^0$&
$\Lambda^0_-$&$\Xi^-_-$&$\Xi^-_0$\\ \hline
exp.&-0.65
&19.07
&12.04
&-7.14
&9.98
&6.93
&-6.43 \\ \hline 
$O_1$
&0.4104
&0
&-0.2902
&1.01387
&-1.434
&0.5559
&-0.3931
\\ \hline
$O_2$
&0.03344
&0
&-0.02365
&0.08261
&-0.1168
&0.04530
&-0.03203
\\ \hline
$O_3$
&0.02736
&0
&-0.01935
&0.06759
&-0.09559
&0.03706
&-0.02621
\\ \hline
$O_4$
&0.1368
&0
&0.1935
&-0.6759
&-0.4779
&0.1853
&0.2621
\\ \hline
$O_{51}$
&0
&0
&0
&0
&0
&0
&0
\\ \hline
$O_{52}$
&6.199E-4
&0.005641
&0.003550
&-0.003244
&0.004588
&-0.002606
&0.001842
\\ \hline
$O_{53}$
&-6.840E-4
&0
&4.837E-4
&-0.001690
&0.002390
&-9.265E-4
&6.551E-4
\\ \hline
$O_{61}$
&-0.01653
&-0.1504
&-0.09467
&0.08651
&-0.1223
&0.06948
&-0.04913
\\ \hline
$O_{62}$
&0.01824
&0
&-0.01290
&0.04506
&-0.06373
&0.02471
&-0.01747
\\ \hline
\end{tabular}
\end{center}
}


\newpage
\vspace{8mm}
\noindent
{\bf\large Table~4.6}\hspace{3mm}
The amplitudes for each operator, which are derived from the one-loop
calculation with data set 2 and $B_0=1.3914\times10^2\mbox{[MeV]}$.
The first column corresponds to the operator number of the
effective Hamiltonian (\ref{eqn:weak5}) and
(\ref{eqn:operator3}). The second row shows the
observed data.\\
{\footnotesize
\begin{center}
\begin{tabular}{|c|c|c|c|c|c|c|c|}
\multicolumn{8}{l}{{\normalsize S-wave:}}\\ \hline
vertex&$\Sigma_-^-$&$\Sigma_+^+$&$\Sigma^+_0$&$\Lambda_0^0$&
$\Lambda^0_-$&$\Xi^-_-$&$\Xi^-_0$\\ \hline
exp.&1.93
&0.06
&-1.48
&-1.07
&1.47
&-2.04
&-1.54 \\ \hline 
$O_1$
&0.8109
&0.03720
&-0.5395
&-0.7886
&1.124
&-1.212
&0.8587
\\ \hline
$O_2$
&0.06608
&0.003031
&-0.04396
&-0.06425
&0.09157
&-0.09876
&0.06996
\\ \hline
$O_3$
&0.1346
&0.004306
&-0.09025
&-0.06023
&0.08568
&-0.1685
&0.1178
\\ \hline
$O_4$
&0.3988
&-0.08284
&0.4980
&0.4776
&0.3318
&-0.3530
&-0.4974
\\ \hline
$O_{51}$
&0
&0
&0
&0
&0
&0
&0
\\ \hline
$O_{52}$
&-0.02663
&0
&0.01884
&0.1102
&-0.1558
&0.05705
&-0.04033
\\ \hline
$O_{53}$
&-0.06330
&0.002354
&0.04215
&0.03044
&-0.04338
&0.06150
&-0.04172
\\ \hline
$O_{61}$
&0.7101
&-0.0002238
&-0.5023
&-2.938
&4.156
&-1.521
&1.075
\\ \hline
$O_{62}$
&1.688
&-0.06277
&-1.124
&-0.8116
&1.157
&-1.640
&1.113
\\ \hline
\end{tabular}
\end{center}
}
{\footnotesize
\begin{center}
\begin{tabular}{|c|c|c|c|c|c|c|c|}
\multicolumn{8}{l}{{\normalsize P-wave:}}\\ \hline
vertex&$\Sigma_-^-$&$\Sigma_+^+$&$\Sigma^+_0$&$\Lambda_0^0$&
$\Lambda^0_-$&$\Xi^-_-$&$\Xi^-_0$\\ \hline
exp.&-0.65
&19.07
&12.04
&-7.14
&9.98
&6.93
&-6.43 \\ \hline 
$O_1$
&0.5634
&6.890
&4.473
&-1.549
&2.191
&0.007857
&-0.005556
\\ \hline
$O_2$
&0.05174
&0.5680
&0.3650
&-0.1240
&0.1753
&0.005649
&-0.003995
\\ \hline
$O_3$
&0.06527
&-0.2370
&-0.2137
&0.1996
&-0.2822
&-0.1340
&0.09473
\\ \hline
$O_4$
&1.028
&-0.004212
&1.450
&0.5049
&0.3570
&-0.2686
&-0.3799
\\ \hline
$O_{51}$
&-3.976E-4
&-4.487E-4
&-3.612E-5
&-1.561E-4
&2.207E-4
&-3.415E-4
&2.415E-4
\\ \hline
$O_{52}$
&-0.8817
&-0.7550
&0.08955
&-1.337
&1.892
&-0.8356
&0.5909
\\ \hline
$O_{53}$
&0.9487
&0.3746
&-0.4058
&1.515
&-2.143
&0.8626
&-0.6098
\\ \hline
$O_{61}$
&23.51
&20.13
&-2.388
&35.67
&-50.44
&22.28
&-15.76
\\ \hline
$O_{62}$
&-25.30
&-9.990
&10.82
&-40.39
&57.13
&-23.00
&16.26
\\ \hline
\end{tabular}
\end{center}
}


\newpage
\vspace{8mm}
\noindent
{\bf\large Table~4.7}\hspace{3mm}
The S- and P-wave $\Delta I=3/2$ amplitudes of each decay mode.
The second column shows the experimentally obtained ratio of
$\Delta I=3/2$ and $\Delta I=1/2$ amplitudes. The third column
represents the $\Delta I=3/2$ amplitudes which are calculated with
the total decay amplitude and the ratio of the second column. The
fourth and fifth column show the $\Delta I=3/2$ amplitude obtained by 
the effective weak Hamiltonian (\ref{eqn:weak5}).\\

\begin{center}
\begin{tabular}{|c|c|c|c|c|}
\multicolumn{5}{l}{S-wave}\\ \hline
decay mode & $A^{3/2}/A^{1/2}$ & $A^{3/2}_{exp.}$ & $A^{3/2}_{tree}$ &
$A^{3/2}_{tree+loop}$ \\ \hline
$\Sigma^-_-$&
$-0.061\pm 0.024$&
-0.1254&
0.06909&
0.4679
\\ \hline
$\Sigma^+_+$&
$-0.061\pm 0.024$&
-0.003898&
0.0&
-0.08284
\\ \hline
$\Sigma^+_0$&
$-0.061\pm 0.024$&
0.09614&
0.09770&
0.5957
\\ \hline
$\Lambda^0_0$&
$0.027\pm 0.008$&
-0.02813&
0.08735&
0.5650
\\ \hline
$\Lambda^0_-$&
$0.027\pm 0.008$&
0.03865&
0.06176&
0.3936
\\ \hline
$\Xi^-_-$&
$-0.038\pm 0.014$&
0.08058&
-0.06978&
-0.4228
\\ \hline
$\Xi^0_0$
&$-0.038\pm 0.014$&
-0.06083&
-0.09868&
-0.5961
\\ \hline
\end{tabular}
\end{center}

\begin{center}
\begin{tabular}{|c|c|c|c|c|}
\multicolumn{5}{l}{P-wave}\\ \hline
decay mode & $B^{3/2}/B^{1/2}$ & $B^{3/2}_{exp.}$ & $B^{3/2}_{tree}$ &
$B^{3/2}_{tree+loop}$ \\ \hline
$\Sigma^-_-$&
$-0.074\pm 0.027$&
0.05194&
0.1368&
1.165
\\ \hline
$\Sigma^+_+$&
$-0.074\pm 0.027$&
-1.524&
0.0&
-0.004212
\\ \hline
$\Sigma^+_0$&
$-0.074\pm 0.027$&
-0.9622&
0.1935&
1.644
\\ \hline
$\Lambda^0_0$&
$0.030\pm 0.037$&
-0.2080&
-0.6759&
-0.1710
\\ \hline
$\Lambda^0_-$&
$0.030\pm 0.037$&
0.2907&
-0.4779&
-0.1209
\\ \hline
$\Xi^-_-$&
$-0.17\pm 0.09$&
-1.419&
0.1853&
-0.0833
\\ \hline
$\Xi^0_0$
&$-0.17\pm 0.09$&
1.317&
0.2621&
-0.1178
\\ \hline
\end{tabular}
\end{center}


\newpage
\setcounter{section}{3}
\setcounter{figure}{0}
\begin{figure}
    \centerline{\epsfxsize=130mm \epsfbox{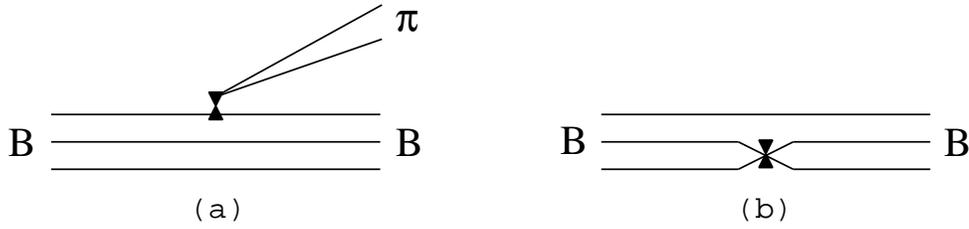}}
    \caption{$\Delta S=1$ hyperon non-leptonic weak decay of quark
             currents.
             (a) corresponds to the weak interaction between two
             hadronic currents which are color singlet.
             (b) corresponds to the weak interaction inside the hyperon.} 
    \label{fig:quarks}
\end{figure}
\setcounter{section}{4}
\setcounter{figure}{0}
\begin{figure}
    \centerline{\epsfxsize=130mm \epsfbox{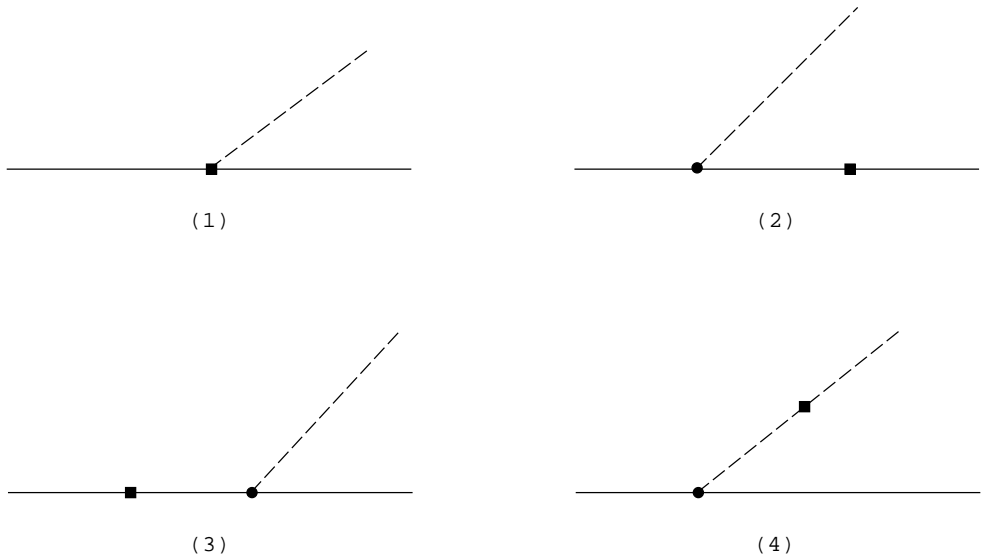}}
    \caption{Tree graph for hyperon non-leptonic decay.
Single solid lines represent octet baryon fields and dots lines
represent meson fields.
The solid square denotes $\Delta S=1$ weak interaction vertices and
the solid dot denotes strong interaction vertices.}
  \label{fig:tree}
\end{figure}
\begin{figure}
   \centerline{\epsfxsize=115mm \epsfbox{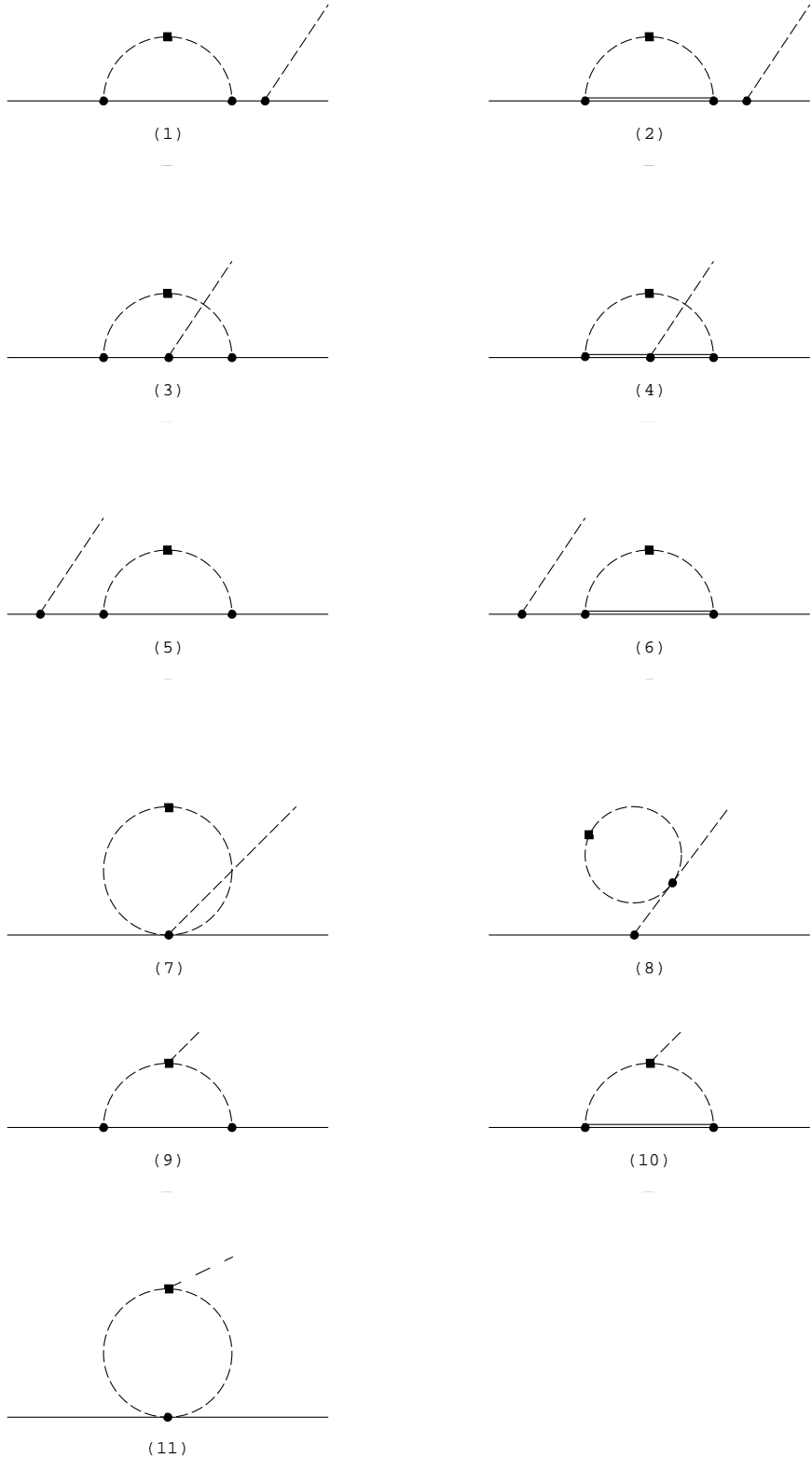}}
    \caption{One loop diagrams for hyperon non-leptonic weak decay.
Single solid lines represent octet baryon fields, double solid lines
represents decuplet fields and dots lines denote meson fields.
the solid square denotes $\Delta S=1$ weak interaction vertices and
the solid dot denotes strong interaction vertices.}
  \label{fig:loop}
\end{figure}
\newpage
\setcounter{figure}{1}
  \begin{figure}
    \centerline{\epsfxsize=115mm \epsfbox{loop2.eps}}
    \caption{(continued)}
  \end{figure}
\newpage
\setcounter{figure}{1}
  \begin{figure}
    \centerline{\epsfxsize=115mm \epsfbox{loop3.eps}}
    \caption{(continued)}
  \end{figure}
\newpage
\setcounter{figure}{1}
  \begin{figure}
    \centerline{\epsfxsize=115mm \epsfbox{loop4.eps}}
    \caption{(continued)}
  \end{figure}
\newpage
\setcounter{figure}{1}
  \begin{figure}
    \centerline{\epsfxsize=115mm \epsfbox{loop5.eps}}
    \caption{(continued)}
\end{figure}
\begin{figure}
  \centerline{\epsfxsize=120mm \epsfbox{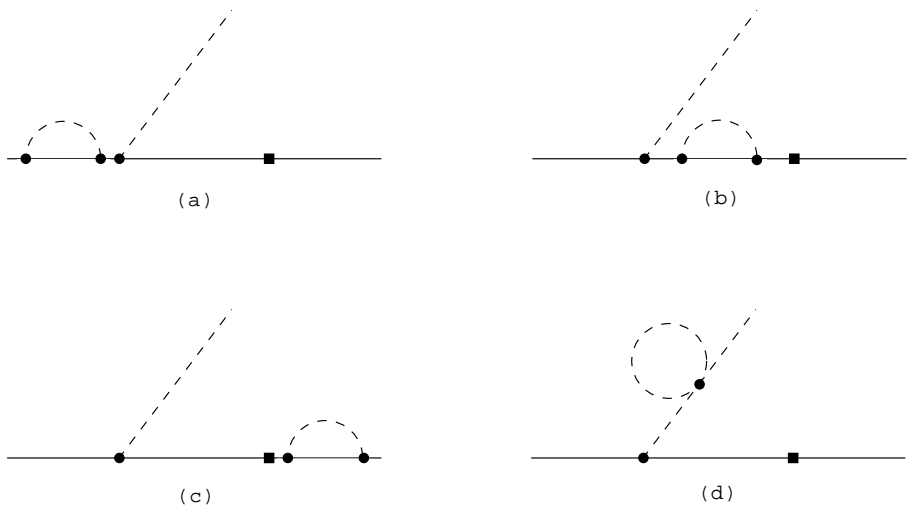}}
  \caption{Examples of the wave function renormalization applied to
           the tree diagram in Figure~4.2 (2). In this study, it is
           considered that the internal baryon state is the decuplet
           baryon.}
  \label{fig:treerenorm}
\end{figure}


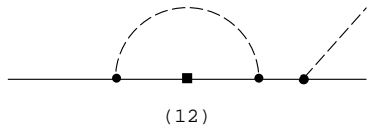
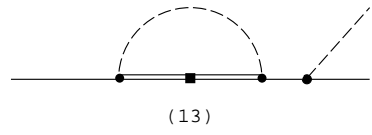
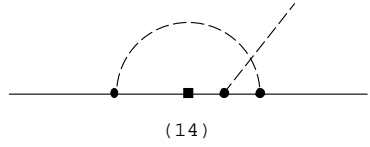
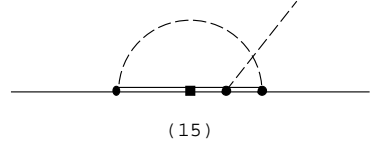
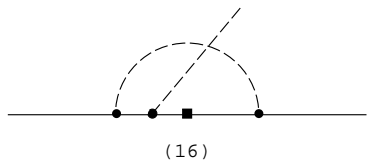
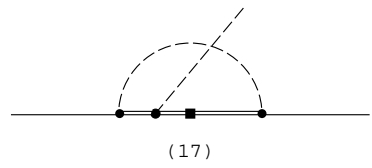
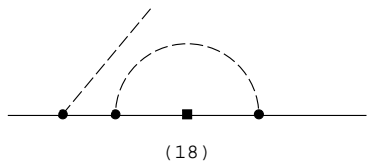
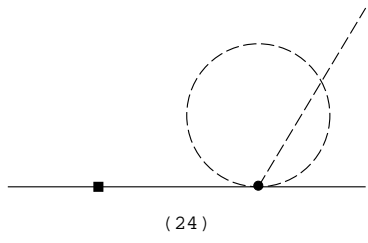
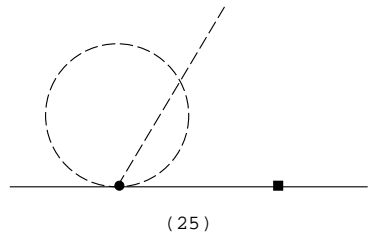
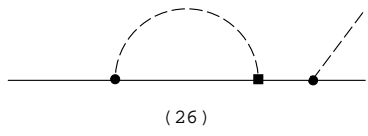
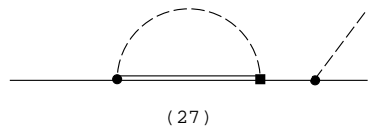
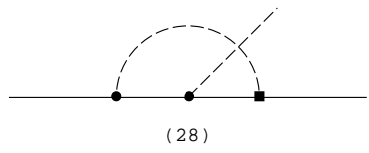
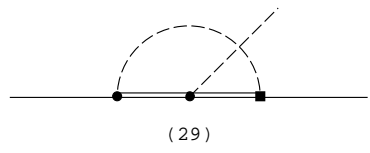
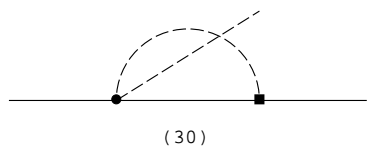
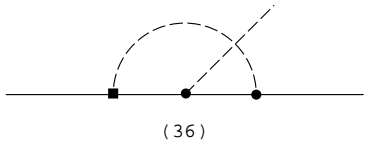
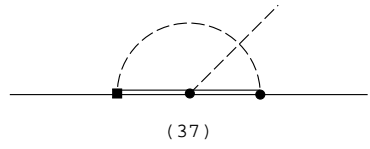
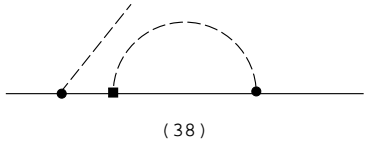
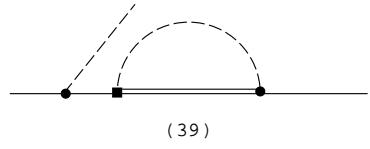
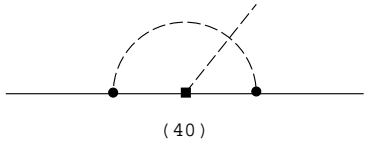
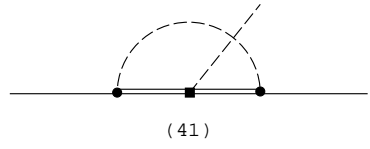
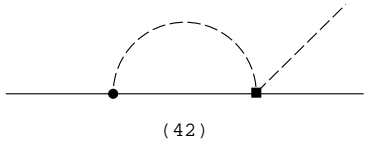
\end{document}